\documentclass[aps,prb,twocolumn,showpacs]{revtex4-1}
\usepackage{graphicx}
\usepackage{amsfonts}
\usepackage{amssymb}
\usepackage{stmaryrd}
\usepackage{wasysym}
\usepackage{pifont}
\usepackage[usenames,dvipsnames]{xcolor}
\usepackage{color,hyperref}

\definecolor{green4}{RGB}{0,139,0}
\definecolor{orange}{RGB}{255,165,0}
\definecolor{grey}{RGB}{150,150,150}
\definecolor{red}{RGB}{255,0,0}
\definecolor{blue}{RGB}{0,0,255}

\begin{document}

\title{The influence of spin fluctuations on the thermal conductivity in superconducting Ba(Fe$_{1-x}$Co$_x$)$_2$As$_2$}

\author{Andrew F. May}
\author{Michael A. McGuire}
\author{Jonathan E. Mitchell}
\author{Athena S. Sefat}
\author{Brian C. Sales}
\affiliation{Materials Science and Technology Division, Oak Ridge National Laboratory, Oak Ridge, Tennessee 37831}

\begin{abstract}
The thermal conductivity of electron-doped Ba(Fe$_{1-x}$Co$_x$)$_2$As$_2$ single crystals is investigated below 200\,K, with an emphasis on the behavior near the magnetic and superconducting ($T_c$) transition temperatures.  An enhancement of the in-plane thermal conductivity $\kappa_{ab}$ is observed below $T_c$ for all samples, with the greatest enhancement observed near optimal doping.  The observed trends are consistent with the scattering of heat carriers by low-energy magnetic excitations.  Upon entering the superconducting state, the formation of a spin-gap leads to reduced scattering and an enhancement in $\kappa(T)$.  Similarly, an enhancement of $\kappa$ is observed for polycrystalline BaFe$_2$As$_2$ below the magnetic transition, and qualitative differences in $\kappa(T)$ between single crystalline and polycrystalline BaFe$_2$As$_2$ are utilized to discuss anisotropic scattering.  This study highlights how measuring $\kappa$ near $T_c$ in novel superconductors can be useful as a means to probe the potential role of spin fluctuations.
\end{abstract}

\maketitle

This article has been published in Physical Review B and can be found at \href{http://link.aps.org/doi/10.1103/PhysRevB.88.064502}{http://link.aps.org/doi/10.1103/PhysRevB.88.064502}.

\section{Introduction}

In unconventional superconductors, which are characterized by a non-phonon mediated pairing mechanism, the thermal conductivity $\kappa$ typically increases upon cooling through the superconducting transition temperature $T_c$.\cite{Ausloos1999,Yu1992,Pogorelov1995,ZhangPRL2001,Movshovich2001}  This enhancement in $\kappa$ below $T_c$ has been observed in the iron-based superconductors.\cite{Machida2011,Checkelsky2012,Nakajima2012}    On the contrary, $\kappa$ in conventional superconductors usually decreases upon cooling through $T_c$ due to the loss of the electronic contribution.\cite{Bardeen1959,Cody1964}  This trend suggest that examining $\kappa(T)$ in the vicinity of $T_c$ can provide insight into the nature of the superconducting pairing mechanism, which can be especially useful when probing the behavior of newly discovered superconductors. 

Of the various iron pnictides, those derived from BaFe$_2$As$_2$ have provided model systems for studying the basic physics in these superconductors.\cite{Sefat2008}    BaFe$_2$As$_2$ is metallic and undergoes a coupled structural (tetragonal to orthorhombic) and magnetic (paramagnetic to antiferromagnetic) transition upon cooling below $\sim$140\,K.\cite{Rotter2008A,Kim2011b} The antiferromagnetic (AFM) transition is associated with a commensurate spin density wave (SDW), with electron-hole nesting vector \textbf{Q}=$(\frac{1}{2}\frac{1}{2}0)$ (tetragonal notation).\cite{Kitagawa2008,NingPRL2010,PrattPRL2011,SinghReview2009}  The magnetic excitation spectrum in the ordered state, as observed via inelastic neutron scattering, is characterized by spin-wave excitations that possess a gap of approximately 10\,meV.\cite{MatanPRB2009,Qureshi2012}  Many detailed studies and review articles discuss the doping dependence of physical properties and magnetism in BaFe$_2$As$_2$.\cite{Sefat2008,NiPRB2008,ChuPRB2009,Lester2009,CanfieldPRB2009,Mun2009,NandiPRL2010,FernandesPRB2010,Mandrus2010,JohnstonReview2010,PaglioneGreen2010,LumsdenChristianson2010}   

Electron doping via cobalt substitution in Ba(Fe$_{1-x}$Co$_x$)$_2$As$_2$ suppresses the structural and magnetic transitions,\cite{Sefat2008,NiPRB2008,Lester2009,Albenque2009,NandiPRL2010} and similar phase diagrams evolve with hole-doping on the Ba site or isoelectronic substitution on the As site.\cite{Rotter2008B,ChenEPL2009,Jiang2009,PaglioneGreen2010} Superconductivity exists for $0.03 \lesssim x \lesssim 0.13$ and even coexists with AFM order for $0.03 \lesssim x \lesssim 0.06$.\cite{ChristiansonPRL2009,PrattPRL2009,PrattPRL2011}  With increasing Co concentration, the magnetic transition occurs at a temperature $T_{SDW}$ that is lower than that of the structural transition $T_{O}$.  At optimal doping ($x \sim 0.06-0.07$), the magnetic and structural transitions are entirely suppressed and superconductivity emerges at $T_{c,max}$$\approx$24\,K from the paramagnetic, tetragonal state.  These materials are known to have a large magnetoelastic coupling,\cite{MazinPRB2008,SinghReview2009} and in underdoped materials both the orthorhombic distortion and the magnetic order parameter decrease upon cooling below $T_c$.\cite{ChristiansonPRL2009,FernandesPRB2010,NandiPRL2010}  For nearly-optimally doped compositions, the tetragonal structure even re-emerges below $T_c$.\cite{NandiPRL2010}  

Upon cooling below $T_c$, a resonance and gap form in the magnetic spectrum at the wave vector associated with the SDW in the parent compound.\cite{LumsdenPRL2009,Inosov2009} Interestingly, the energy of the resonance ($E_{r}$) is found to scale with $T_c$ in a manner very similar to the behavior observed in the cuprates ($E_{r} \approx 5kT_c$).\cite{Hufner2008,LumsdenChristianson2010} 

With cobalt doping the ordered moment decreases,\cite{FernandesPRB2010} and the spin-gap decreases and broadens as the spin-wave dispersions transform to a magnetic excitation spectrum similar to that in the paramagnetic state (characterized by strong fluctuations).\cite{Lester2010,TuckerPRB2012} For the superconducting compositions, the spin-gap is largest at optimal doping, and decreases with increasing (or decreasing) cobalt concentration.   Magnetic fluctuations are not present in the non-superconducting, overdoped compositions ($x \geq 0.15$) due to the disappearance of the hole pocket and the associated loss of an available nesting vector.\cite{MatanPRB2010,NingPRL2010}

Here, we show that $\kappa$ increases below $T_c$ in Ba(Fe$_{1-x}$Co$_x$)$_2$As$_2$ crystals with compositions where a spin-gap forms below $T_c$.  The spin-gap prohibits the formation of low-energy magnetic excitations that could otherwise scatter heat carriers, such as quasiparticles or phonons.  To demonstrate this behavior, $\kappa(T)$ is examined near $T_c$ in Ba(Fe$_{1-x}$Co$_x$)$_2$As$_2$ single crystals with compositions ranging from $x=0$ to $x\sim0.2$.  Importantly, a \textit{slightly} underdoped ($T_c\approx$21\,K) crystal is characterized, which allows the potential role of nematic fluctuations to be addressed.  In this case, $\kappa$ does indeed increase below $T_c$ even though nematic fluctuations are frozen out well above $T_c$, revealing that nematic (or structural) fluctuations are most likely not a dominant source for scattering.

\begin{table}%
\begin{tabular}{cccccc}
\hline
$x$   & symbol &  $T_{c,onset}$ & $T_{c,50\%}$ & $T_{O}$ & $T_{SDW}$ \\
-    & - & K  & K & K & K \\
\hline
0 &        {\color{black}$\blacktriangleleft$}        & -       & -       & 137    & 137   \\ 
0.043 &   {\color{green4} \ding{117}}                & 17.7    & 16.6    & 70  &  55     \\  
0.049 &   {\color{orange}\ding{116}}               & 21.6    & 21.1    & 53   & 35     \\  
0.075 &   {\color{red}\ding{115}}              & 24.8    & 24.4    & -       & -  \\ 
0.11  &   {\color{blue}\ding{108}}                & 14.8    & 13.9    & -    & -     \\ 
0.20  &   {\color{grey}$\blacktriangleright$}       & $<$2\,K & $<$2\,K & -  & -       \\ 
\hline
\end{tabular}
\caption{Characteristic properties of the Ba(Fe$_{1-x}$Co$_x$)$_2$As$_2$ crystals with $x$ obtained from energy dispersive spectroscopy (all standard deviations $<$0.003) and transition temperatures obtained from electrical resistivity data; superconducting transitions $T_c$ are given for both the onset and 50\% resistive change criteria.}
\label{tab}
\end{table}

\section{Experimental Details}
Single crystals of Ba(Fe$_{1-x}$Co$_x$)$_2$As$_2$ were grown from an FeAs flux using melts of nominal composition Ba(Fe$_{1-x}$Co$_x$)$_5$As$_5$.  Dendritic Ba was combined with FeAs and CoAs in an Al$_2$O$_3$ crucible and sealed in a silica ampoule with approximately 1/5\,atm of argon. The mixtures were heated to between 1180$^{\circ}$C and 1220$^{\circ}$C, followed by cooling at 2$^{\circ}$/h to approximately 1090$^{\circ}$C, at which temperature the samples were taken from the furnace and excess FeAs/CoAs was removed by centrifugation.  

Cobalt concentrations ($x$) were determined using the relative amounts of Fe and Co obtained from energy dispersive spectroscopy (EDS) in a Bruker Quantax 70 EDS system on a Hitachi TM-3000 microscope.  Approximately 15 different spots were examined on each crystal (each measurement encompassing a diameter of 200-400$\mu$m).  An average $x$ based on all measurements is reported in Table \ref{tab}.  The standard deviations of these $x$ values are $<$ 0.003 for all measurements, demonstrating the homogeneity of the samples.  It is possible there is a systematic error in these values due to the measurement technique, and an equally valid characterization is through the observation of anomalies in the electrical resistivity associated with the various phase transitions, and comparison with published phase diagrams.\cite{NiPRB2008,ChuPRB2009,Lester2009,NandiPRL2010,PaglioneGreen2010}  The derivative of the resistivity is utilized to obtain $T_{O}$ and $T_{SDW}$ in accord with References \citenum{NiPRB2008,ChuPRB2009}.  The superconducting transition temperature $T_c$ is obtained using both the 50\% resistive change ($T_{c,50\%}$) and onset ($T_{c,onset}$, maximal slope method) approaches.  These values are presented in Table \ref{tab}, along with the sample identification symbols utilized in this manuscript.  In general, these results agree with the literature and small variation between our results and those from the literature could be caused by differences in sample preparation\cite{Gofryk2011} or measurement errors.

Ba(Fe$_{1-x}$Co$_x$)$_2$As$_2$ crystals grow as plates with large facets characterized by the tetragonal [001] normal. As such, transport measurements within the $\textit{ab}$-plane are the easiest and most reliable ones to perform.  Measurements perpendicular to this face generally require a less accurate two-point configuration and are prone to failure and errors due to delamination of the crystals.  Large crystals were grown for thermal transport measurements, with thicknesses ranging from $\approx$0.15\,mm to 0.45\,mm, and lengths between heat source and sink were typically $\approx$4\,mm or more.  Thermal transport measurements were performed in a Quantum Design Physical Property Measurement System using the Thermal Transport Option.  Silver epoxy (H20E Epo-Tek) was used for electrical, thermal, and mechanical contacts in a standard four-point configuration.  The error on the thermal conductivity may be approximately 10\%, due primarily to geometric concerns.  Error in the absolute values obtained do not influence the trends observed, which are primarily based on relative changes in $\kappa$.  The AC Transport Option was utilized to obtain the electrical resistivity in underdoped compositions for the determination of $T_{O}$ and $T_{SDW}$ from analysis of d$\rho$/d$T$.\cite{NiPRB2008,ChuPRB2009}

Polycrystalline samples of nominal compositions Ba$_{1.05}$Fe$_2$As$_2$ and Ba$_{1.05}$(Fe$_{0.95}$Co$_{0.05}$)$_2$As$_2$ were synthesized to probe the role of anisotropy in the parent and underdoped compositions.  Elemental Ba (dendritic) was reacted with FeAs and CoAs in alumina crucibles, which were sealed within evacuated silica ampoules.  The mixtures were heated to 850\,$^{\circ}$C, followed by a 10\,h soak prior to homogenization and subsequent reaction at 900\,$^{\circ}$C for 50\,h.  The products were ground once again, pressed into pellets at approximately 40,000\,psi, and sintered at 900\,$^{\circ}$C for 20\,h.  This resulted in $\sim$80\% of the theoretical density, and the pellets were found to be phase pure by powder x-ray diffraction.  Powders were handled in a helium glove box prior to sintering.

\begin{figure}[!ht]
\includegraphics [width=3.2in] {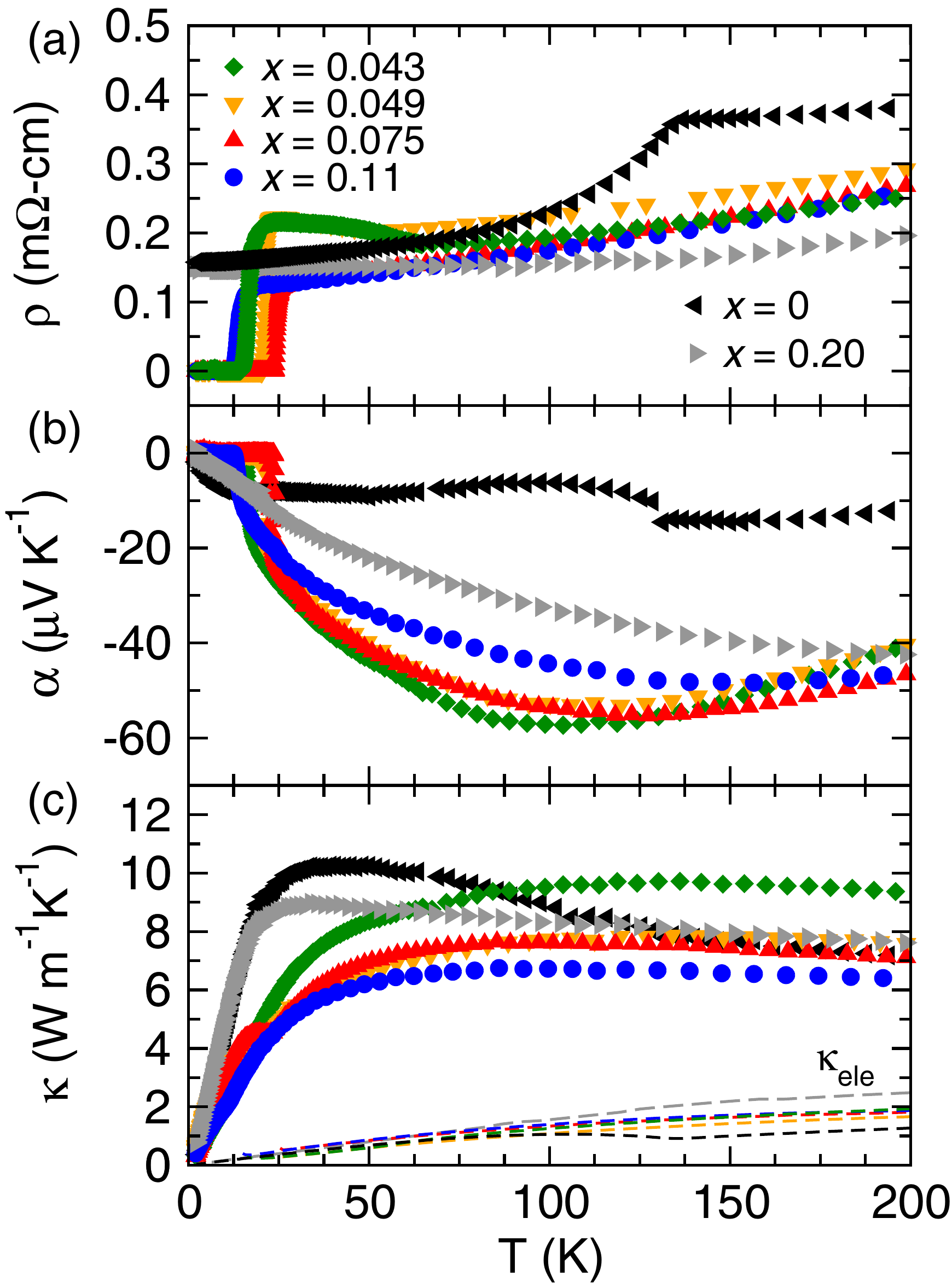}
\caption {The in-plane (a) electrical resistivity, (b) Seebeck coefficient, and (c) thermal conductivity of Ba(Fe$_{1-x}$Co$_x$)$_2$As$_2$ single crystals along with the estimated electronic contribution $\kappa_{\mathrm{ele}}$.  The low temperature data are highlighted in Fig.\,\ref{LowT}.}
\label{TTO}
\end{figure}

A single band Wiedemann-Franz relationship was used to estimate the electronic contribution to the thermal conductivity $\kappa_{\mathrm{ele}}$ = $L\sigma T$, where the Lorenz number $L$ is taken as the degenerate limit (2.44$\times$10$^{-8}$W$\Omega$K$^{-2}$).  The lattice contribution was calculated by $\kappa_{\mathrm{lat}}$=$\kappa$-$\kappa_{\mathrm{ele}}$.  The assumed value of $L$ is reasonable considering the metallic conductivity of these materials, though the multi-band nature of these compounds reduces the validity of this simplified approach (particularly at the lower doping levels).

\section{Results}

The coupled structural and magnetic transitions in BaFe$_2$As$_2$ near 137\,K are easily observed in the electrical resistivity ($\rho$) data shown in Fig.\,\ref{TTO}a.  A corresponding feature is also observed in the Seebeck coefficient ($\alpha$, Fig.\,\ref{TTO}b).  As observed, $\rho$ and $\alpha$ change systematically with increasing cobalt content in Ba(Fe$_{1-x}$Co$_x$)$_2$As$_2$, and these results are consistent with the literature.\cite{Mun2009}

\begin{figure}[!ht]
\includegraphics [width=3.2in] {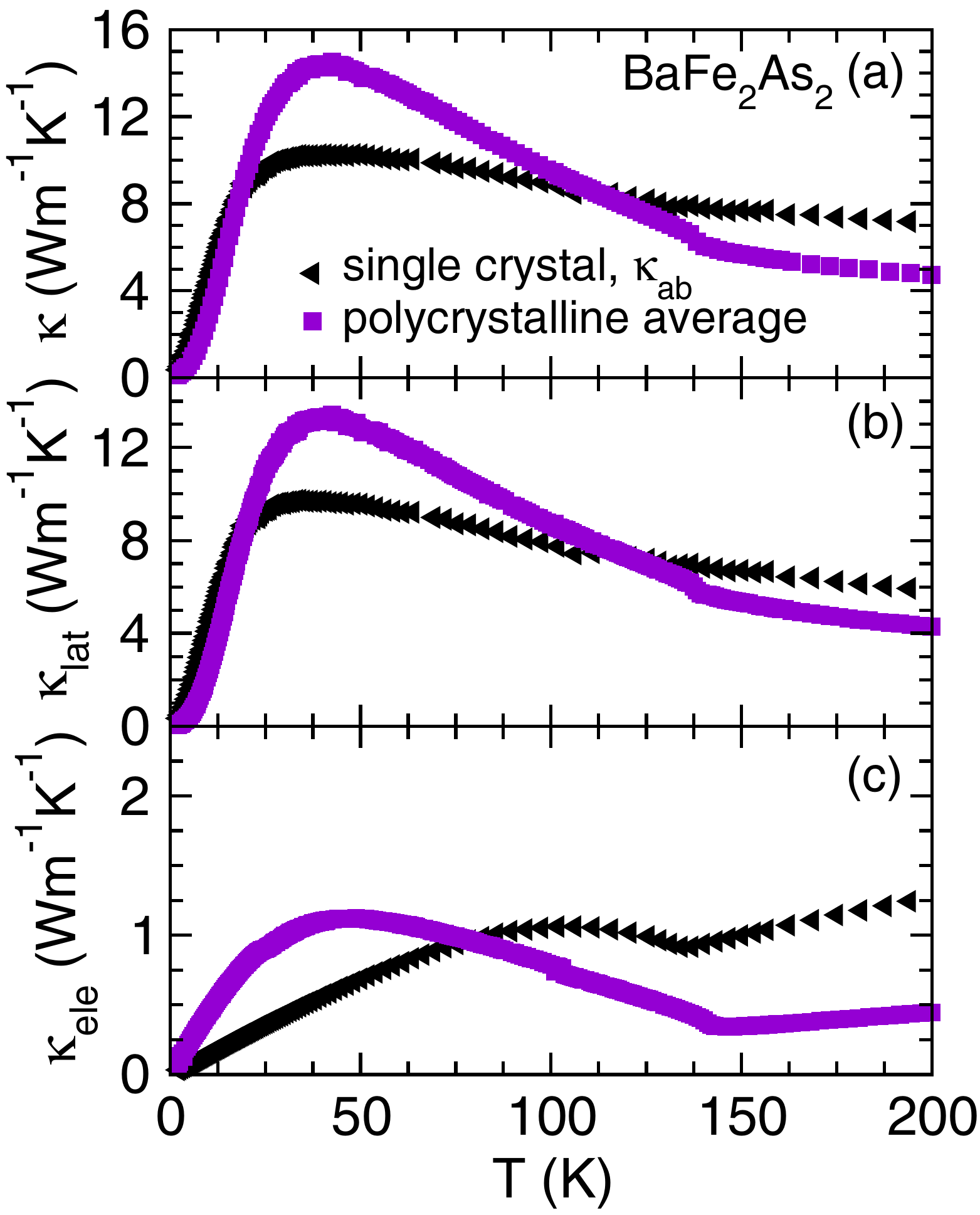}
\caption{A comparison of the thermal transport data for polycrystalline and single crystalline ($\kappa_{ab}$) BaFe$_2$As$_2$, with the (a) total, (b) lattice, and (c) electronic components of the thermal conductivity shown.}
\label{KappaPoly}
\end{figure}

Figure \ref{TTO}c shows the measured thermal conductivity of Ba(Fe$_{1-x}$Co$_x$)$_2$As$_2$ single crystals (in-plane $\kappa_{ab}$), as well as the estimated electronic contribution $\kappa_{\mathrm{ele}}$.  As inferred from the small values of $\kappa_{\mathrm{ele}}$, the lattice contribution is a significant portion of the total thermal conductivity.  There is no significant change in the temperature dependence of $\kappa_{ab}$ in the parent BaFe$_2$As$_2$ at the structural/magnetic transition.  Figure \ref{KappaPoly} compares $\kappa_{ab}$ in a single crystal to $\kappa$ in polycrystalline BaFe$_2$As$_2$, and a clear anomaly is observed in $\kappa(T)$ for the polycrystalline sample at the structural/magnetic transition.  We note that no anomalies were observed in $\kappa(T)$ for polycrystalline Ba(Fe$_{0.95}$Co$_{0.05}$)$_2$As$_2$, which possessed $T_{O}$$\approx$80\,K and $T_{c,50\%}$=21.5\,K (not shown).

\begin{figure}[!ht]
\includegraphics [width=3.2in] {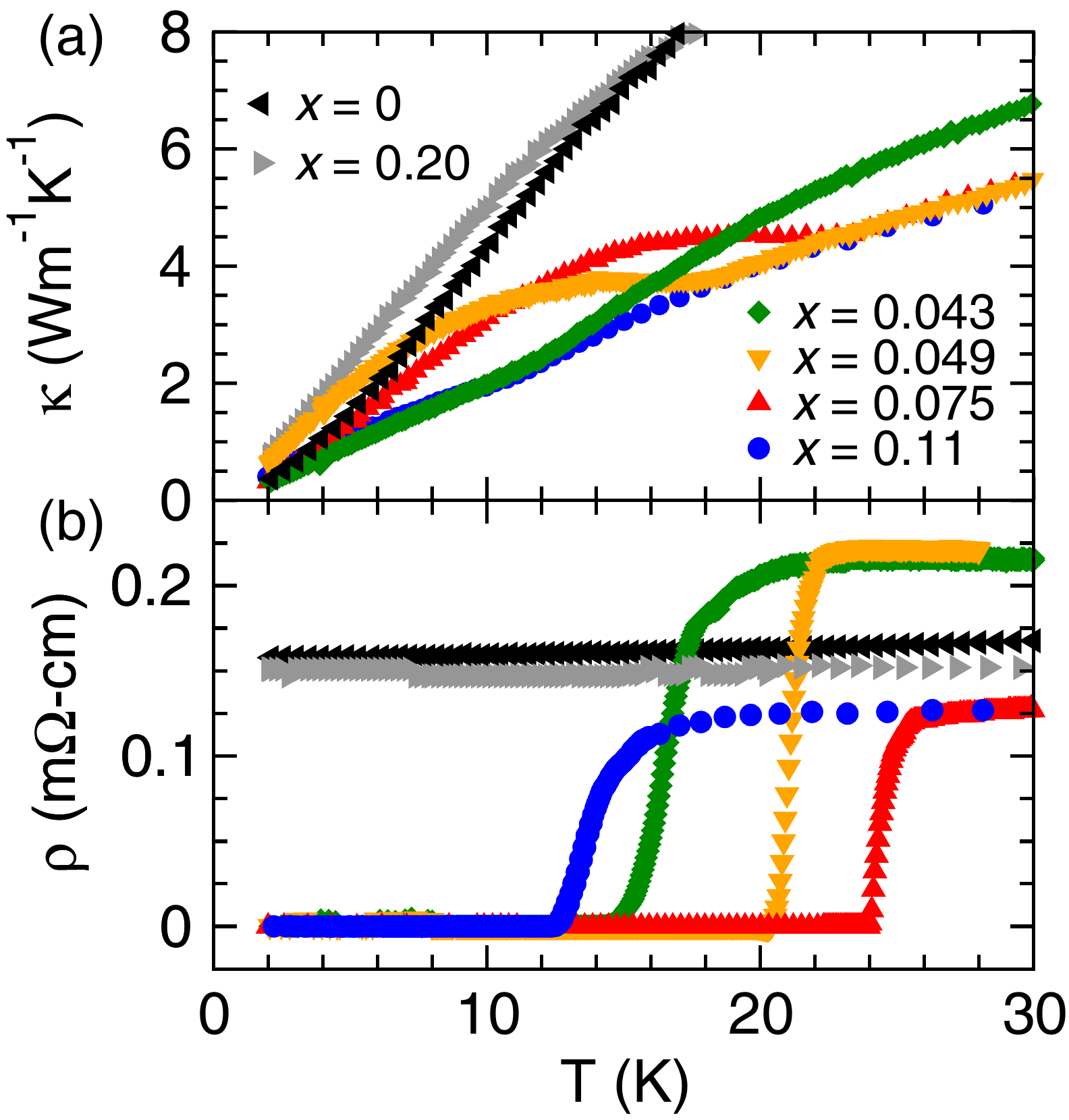}
\caption {The in-plane thermal conductivity and electrical resistivity of Ba(Fe$_{1-x}$Co$_x$)$_2$As$_2$ single crystals at low temperatures highlighting the behavior around $T_c$.}
\label{LowT}
\end{figure}

Figure \ref{LowT} emphasizes the behavior of $\kappa$ and $\rho$ near $T_c$.  As shown, $\kappa$ clearly increases upon cooling below $T_c$ for the nearly optimally doped sample ($x$=0.075) with $T_{c,onset} = 24.8$\,K, as well as for the slightly underdoped sample with $x$=0.049 and $T_{c,onset}=21.6$\,K.  A small increase in $\kappa$ is observed for the overdoped $x$=0.11 below $T_{c,onset}$=14.8\,K, and there is a small change in the temperature dependence of $\kappa$ below $T_c$ for the underdoped sample with $x$=0.043 and $T_{c,onset}$=17.7\,K.  We emphasize that the underdoped sample $x$=0.049 is a special case, because this composition undergoes the structural/magnetic transitions well above $T_c$ but $\kappa$ still increases sharply below $T_c$.  

\section{Discussion}

We first consider the decrease in the electrical resistivity $\rho$ below $T_{SDW}$ in the parent BaFe$_2$As$_2$.  This behavior is relatively common across magnetic transitions, where spin fluctuations above the transition temperature cause an increase in charge carrier scattering and thus a larger electrical resistivity.  Similarly, spin fluctuations can scatter heat carriers (electrons, phonons, spin waves, quasiparticles).  For instance, the interactions of spin fluctuations and phonons have been nicely shown in YMnO$_3$ and  Y$_3$Fe$_5$O$_{12}$,\cite{SharmaPRL2004,Bhandari1966}  and may explain the anisotropic $\kappa$ observed in CrSb$_2$.\cite{SalesCrSb2012}

As shown in Fig.\,\ref{KappaPoly}, the electronic contribution $\kappa_{\mathrm{ele}}$ and the lattice thermal conductivity $\kappa_{\mathrm{lat}}$ increase upon cooling below $T_{SDW}$ in polycrystalline BaFe$_2$As$_2$.  The increase in $\kappa_{\mathrm{ele}}$ is easily understood in terms of the change in $\rho(T)$ due to reduced scattering below $T_{SDW}$. For $\kappa_{\mathrm{lat}}$, we observe a much weaker temperature dependence above $T_{SDW}$, which indicates the presence of an additional scattering mechanism above $T_{SDW}$.   This change in $\kappa(T)$ was also observed in polycrystalline samples of undoped $Ln$FeAsO ($Ln=$La to Nd) at the combined magnetic/structural transition.\cite{McGuire_LaFeAsO,McGuire2009LnFeAsO}  These changes in scattering can be understood in terms of a reduction in spin or structural fluctuations below the phase transition.

The increase in $\kappa$ upon cooling through $T_{SDW}$ in polycrystalline BaFe$_2$As$_2$ but not in $\kappa_{ab}$ of a single crystal reveals an anisotropy that suggests magnetic excitations above $T_{SDW}$ strongly scatter heat carriers with momentum along the $c$-axis in BaFe$_2$As$_2$.  In single crystalline BaFe$_2$As$_2$, $\rho_c$/$\rho_{ab}$ is always greater than unity but decreases below $T_{SDW}$.\cite{TanatarPRB2009} Together with the current results, it seems that spin fluctuations do indeed strongly scatter electrons/holes and phonons in BaFe$_2$As$_2$ traveling along $c$.  These observations may be related to anisotropy in the magnetic excitation spectrum. In the paramagnetic state, BaFe$_2$As$_2$ is characterized by uncorrelated out-of-plane spins with a broad magnetic scattering intensity.\cite{MatanPRB2009} Below $T_{SDW}$, the excitations are three-dimensional but anisotropic, with in-plane spin-wave velocities ($v_{ab}$ $\sim$280\,meV\,\AA) larger than along the $c$-axis ($v_c$ $\sim$57\,meV\,\AA).\cite{MatanPRB2009}  In addition, the spin-gap is larger for in-plane excitations ($\approx$19\,meV) than out-of-plane excitations ($\approx$12\,meV).\cite{Qureshi2012}  The evolution of the magnetic excitations and inferred anisotropy with doping is quite interesting, and highlights the sensitivity of $\kappa$ to changes in the magnetic excitation spectrum in these materials.

We now focus on the behavior of $\kappa_{ab}$ in single crystals that display superconductivity.  Perhaps the best way to characterize/identify these samples is through their transition temperatures, which are shown in Table\,\ref{tab}.  Underdoped samples possess an increase in $\rho(T)$ upon cooling through the structural transition at $T_{O}$, while optimal- or over-doped samples do not experience the structural distortion.

One clear trend observed in Fig.\,\ref{TTO} and Fig.\,\ref{LowT} is that all superconducting samples have suppressed $\kappa$ at low $T$ relative to the parent or the heavily-overdoped composition ($x$=0.20).  This is additional evidence that scattering by spin fluctuations is important in the superconducting compositions, because these two non-superconducting compositions do not possess low-energy spin fluctuations at low $T$.  The lack of low-energy spin fluctuations in BaFe$_2$As$_2$ is due to the formation of a $\sim$10\,meV spin-gap below $T_{SDW}$.\cite{MatanPRB2009}  For the heavily-overdoped $x$=0.20 crystal, the magnetic excitation spectrum is fundamentally changed due to the lack of electron/hole nesting at this high electron concentration, which leads to a drastic suppression of spin fluctuations.\cite{MatanPRB2010}

As observed in Fig.\,\ref{LowT}, $\kappa$ clearly increases below $T_c$ in the nearly-optimally doped samples ($x$=0.049 and $x$=0.075).  The change in $\kappa$ is much less significant in underdoped $x$=0.043 and overdoped $x$=0.11, though slight enhancements in $\kappa$ can be inferred through changes in the temperature dependence.

\begin{figure}[!ht]
\includegraphics [width=3.2in] {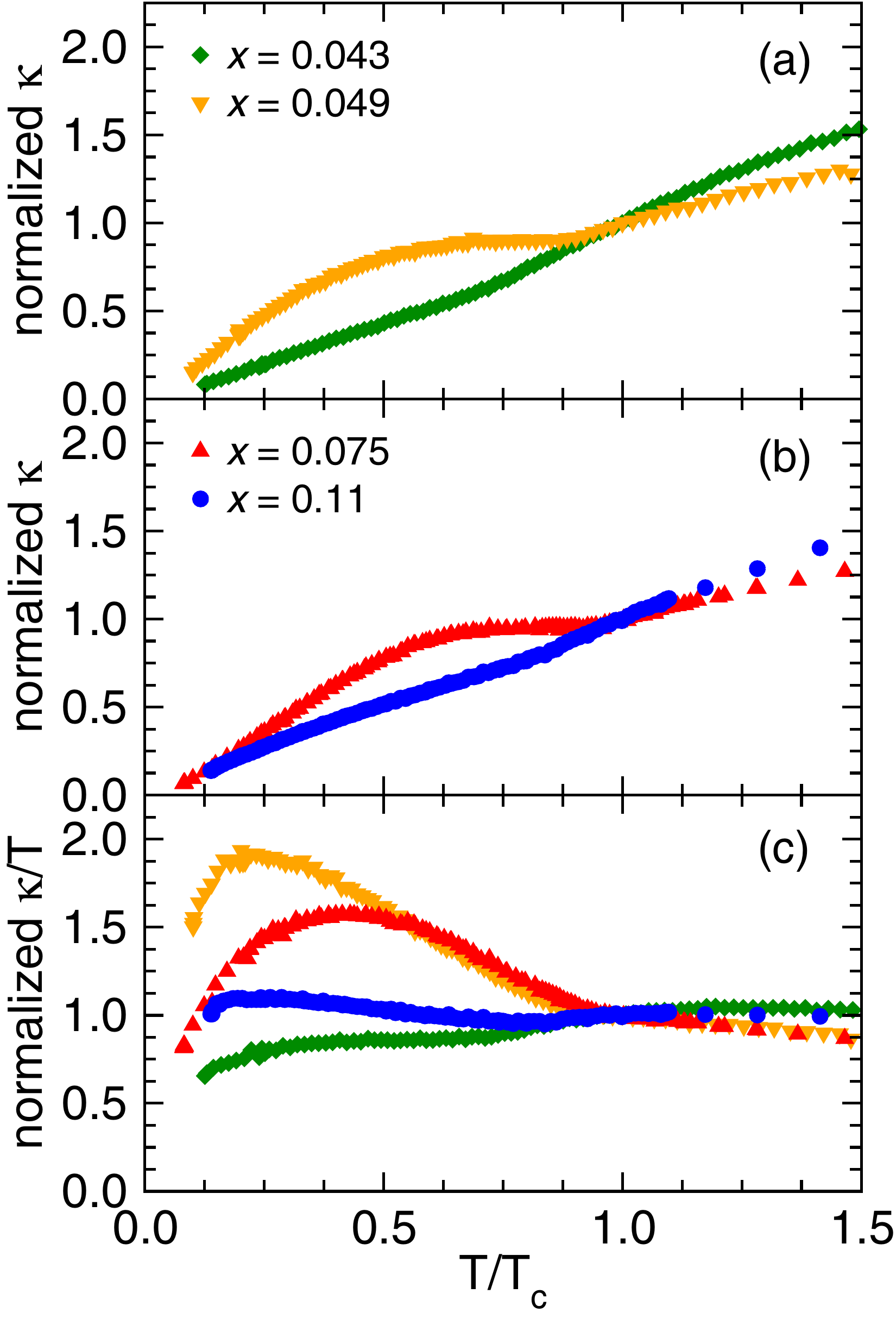}
\caption {(a) The thermal conductivity ($\kappa_{ab}$) of Ba(Fe$_{1-x}$Co$_x$)$_2$As$_2$ normalized to $\kappa_{ab}(T_{c,50\%})$; (b) the in-plane $\kappa/T$ data are normalized to the values of this quantity at $T_{c,50\%}$.}
\label{Normalized}
\end{figure}

\begin{figure}[!ht]
\includegraphics [width=3.2in] {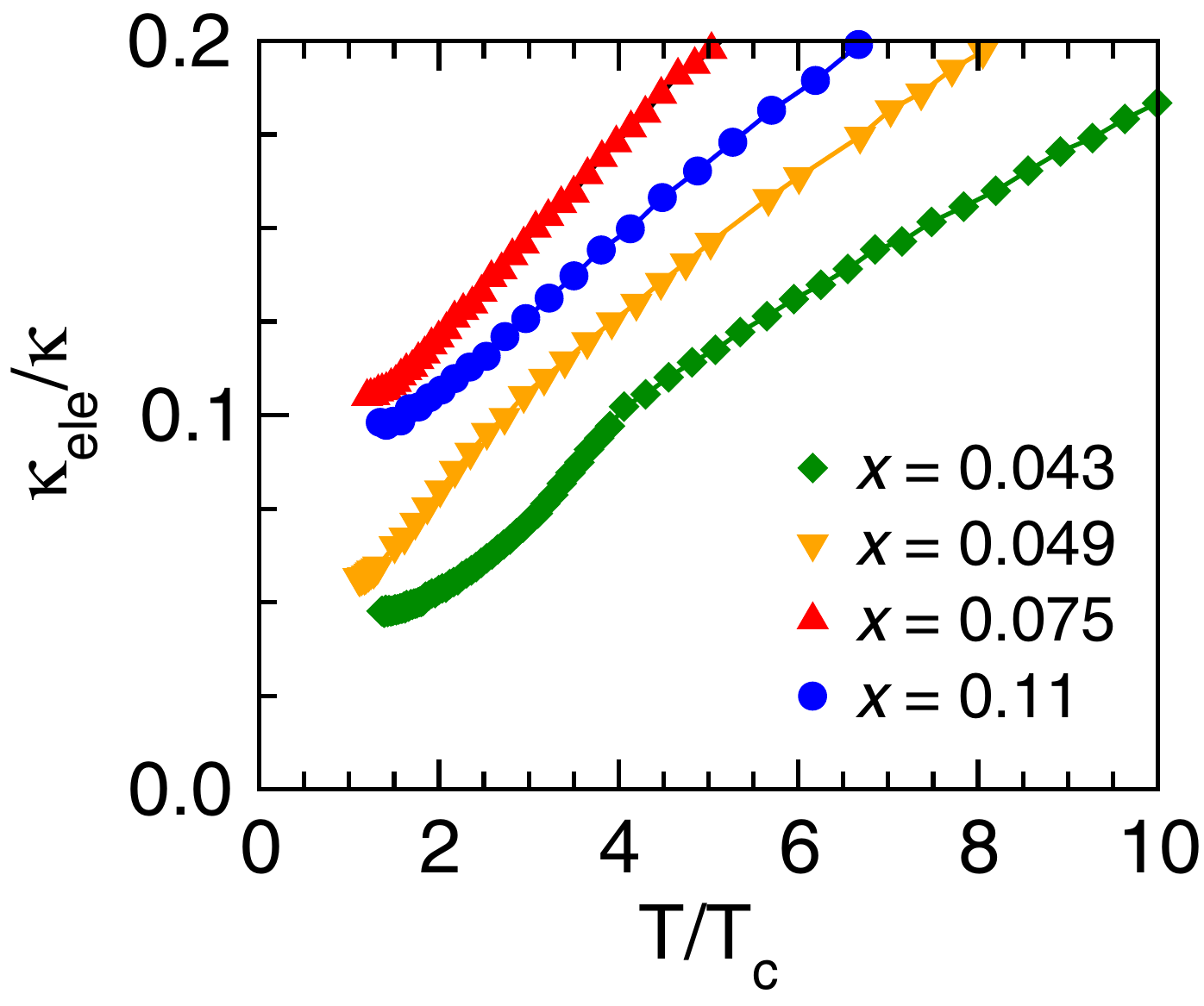}
\caption {The relative importance of the (in-plane) electronic contribution $\kappa_{\mathrm{ele}}$ in Ba(Fe$_{1-x}$Co$_x$)$_2$As$_2$ at low $T$.}
\label{KeKt}
\end{figure}

To highlight the behavior below $T_c$, the data are normalized and plotted in Fig.\,\ref{Normalized}.  A slight increase in $\kappa$ below $T_c$ can be observed for $x$=0.11 in Fig.\,\ref{Normalized}b.  This behavior is more readily observed in the plot of $\kappa/T$ (Fig.\,\ref{Normalized}c), where the relative increase in $\kappa/T$ can be observed for $x$=0.11 below $\approx$0.85$T_c$.  The relative increase in $\kappa/T$ is smaller for $x$=0.043, though a slight increase in $\kappa/T$ can be observed below approximately 0.75$T_c$.  There is clearly a large difference in the behavior of $\kappa(T)$ between $x$=0.049 and $x$=0.043, despite a relatively small change in $T_c$ (or composition).  In summary, all samples show at least a small, relative increase in $\kappa$ as observed through $\kappa/T$ or the temperature dependence of $\kappa$.  The relative increase does not trend with $T_c$, however, as exemplified by the smaller enhancement in $\kappa$ for underdoped $x$=0.043 as opposed to overdoped $x$=0.11.

As shown in Fig.\,\ref{KappaPoly}, both the lattice and electronic components of $\kappa$ can be influenced by spin fluctuations in BaFe$_2$As$_2$ materials. Righi-Leduc measurements (thermal Hall effect) have shown that $\kappa_{\mathrm{ele}}$ increases rapidly below $T_c$ in optimally doped Ba(Fe$_{1-x}$Co$_x$)$_2$As$_2$\cite{Machida2011} and K-doped Ba$_{1-x}$K$_x$Fe$_2$As$_2$,\cite{Checkelsky2012,Carbotte2011} and similar results were shown for the high temperature superconductor YBa$_2$Cu$_3$O$_7$.\cite{ZhangPRL2001} These measurements also reveal a small increase in the lattice component $\kappa_{\mathrm{lat}}$, though the increase in $\kappa_{\mathrm{ele}}$ below $T_c$ is much more significant.\cite{Checkelsky2012}  Therefore, it would be possible for the relative change in $\kappa$ to be significantly suppressed if $\kappa_{\mathrm{ele}}$ were much less than $\kappa_{\mathrm{lat}}$.  As such, the relative contributions of $\kappa_{\mathrm{ele}}$ are plotted in Fig.\,\ref{KeKt}. The ratio of $\kappa_{\mathrm{ele}}$/$\kappa$ is similar for $x$=0.049 and $x$=0.043 near $T_c$, yet the crystal with $x$=0.049 has a much larger $\kappa$ enhancement below $T_c$ than is observed for $x$=0.043.  In addition, $\kappa_{\mathrm{ele}}$/$\kappa$ is larger for $x$=0.11 than for $x$=0.049, though the relative increase in $\kappa$ is much larger for $x$=0.049 compared to $x$=0.11.  The variation of the relative enhancements in $\kappa(T)$ below $T_c$ is therefore not an artifact induced by the relative contributions of $\kappa_{\mathrm{ele}}$.

The experimental data can be explained by relatively simple scattering considerations: Low-energy magnetic excitations scatter heat carriers, and thus the formation of a gap in the magnetic excitation spectrum eliminates a scattering source and results in a relative increase in $\kappa$.  It is important to stress that this line of reasoning is valid regardless of whether or not the dominant heat carriers are electrons or phonons, though in this case the current literature suggests the enhanced $\kappa$ mostly originates in $\kappa_{\mathrm{ele}}$.  For optimally doped and overdoped samples, where magnetic order does not occur, the formation of a superconductivity-induced spin-gap results in an increase in $\kappa(T)$ below $T_c$.  The magnitude of the spin-gap is expected to decrease with increasing cobalt concentration above the optimal doping, as does the strength of the spin fluctuations,\cite{NingPRL2010} which explains the smaller enhancement of $\kappa$ for $x$=0.11 relative to $x$=0.075.  The nature of the spin-gap is more complicated in underdoped Ba(Fe$_{1-x}$Co$_x$)$_2$As$_2$, which become superconducting from a magnetically ordered state and coexistence of the two states occurs for particular ($x$,$T$).

For small cobalt concentrations, the magnetic excitation spectrum in the magnetically ordered state likely remains similar to that in the parent composition, which is characterized by a well-defined spin-gap of $\sim$10\,meV.\cite{MatanPRB2009}  As the cobalt concentration increases, however, the spin-gap is either lost or strongly broadened, and the magnetic spectrum evolves to be similar to that in the paramagnetic phase.\cite{TuckerPRB2012}  Tucker and colleagues studied a crystal with $x=0.047$, $T_{O}\sim$60\,K and $T_{SDW}$=47\,K and found that it did not possess a well-defined spin-gap for $T_c < T < T_{SDW}$.\cite{TuckerPRB2012}  This explains the lack of an increase in $\kappa_{ab}$ at $T_{SDW}$ for $x$=0.043 and $x$=0.049, as well as the smooth $\kappa(T)$ observed in the polycrystalline sample with $x=0.05$.  

In addition to changes in the formation of a spin-gap below $T_{SDW}$, the magnetic excitations become more short-range and two-dimensional with cobalt doping.\cite{LumsdenPRL2009,ChristiansonPRL2009,LiPRB2010}  For compositions that have long-range AFM order, the resonance has a more dispersive behavior along $c$ similar to that of the spin waves in the SDW state.\cite{PrattPRB2010}  The changes in the dimensionality of the magnetic excitation spectrum are manifested in changes in the anisotropy of $\kappa(T)$, which is inferred from differences between $\kappa_{ab}$ in single crystals and $\kappa$ in polycrystalline samples.  In the superconducting samples, relative changes/increases in $\kappa_{ab}(T)$ are observed at $T_c$. This is due to the formation (below $T_c$) of a gap in the magnetic excitation spectrum, which is highly two-dimensional. The excitation spectrum is known to be more three-dimensional for undoped BaFe$_2$As$_2$, though, and $\kappa_{ab}(T)$ is not influenced by the formation of a spin-gap.  In polycrystalline BaFe$_2$As$_2$, however, $\kappa(T)$ is clearly influenced by the SDW/structural transition.  This reveals that the transition to a more two-dimensional magnetic spectrum upon doping leads to greater interaction with heat carriers traveling in the $ab$-plane.  This is one reason the relative change in $\kappa_{ab}$ below $T_c$ is small for underdoped $x$=0.043, while another contributing factor is that the spin-gap is not well-defined and does not change significantly below $T_c$ for underdoped compositions.\cite{ChristiansonPRL2009,TuckerPRB2012}

The potential role of structural/nematic fluctuations warrants discussion.  Local magnetic fluctuations couple to the lattice causing a local orthorhombic distortion (nematic fluctuation).\cite{Fernandes2012}  These fluctuations exist in all compositions for $T$ greater than the structural transition temperature $T_{O}$ or superconducting transition $T_c$, whichever is greater.\cite{Fernandes2012}  The influence of these nematic fluctuations is readily observed through changes in the elastic constants and in-plane resistivity anisotropy.\cite{FernandesPRL2010,ChuScience2010,ChuScience2012} It may therefore seem reasonable that the loss of nematic fluctuations leads to a relative rise in $\kappa$ at $T_c$.  However, nematic fluctuations would be frozen-out at $T_{O}$$\approx$50\,K for $x$=0.049, but a strong increase in $\kappa(T)$ is observed below $T_c$ for this sample.  Furthermore, no changes in $\kappa(T)$ were observed across the structural transition in any underdoped sample.  Therefore, scattering from nematic/structural fluctuations is most likely not responsible for the observed trends in $\kappa(T)$.

In summary, we have examined $\kappa(T)$ in the vicinity of the phase transitions in Ba(Fe$_{1-x}$Co$_{x}$)$_2$As$_2$.  The behavior of $\kappa(T)$ across the magnetic and superconducting transitions can be understood by considering changes in scattering due to the evolution of the magnetic excitation spectrum with composition and temperature.  In nearly-optimally doped or overdoped samples, $\kappa(T)$ increases below $T_c$ due to the formation of a gap in the excitation spectrum.  In underdoped compositions, only a small change in $\kappa$ can be observed below $T_c$ because superconductivity emerges from a magnetically ordered state characterized by a weak spin-gap that does not change significantly at $T_c$.  In addition, the evolving dimensionality of the magnetic excitation spectrum has been revealed through differences between $\kappa_{ab}(T)$ in single crystals and $\kappa(T)$ for polycrystalline materials.  In BaFe$_2$As$_2$, the excitations are three-dimensional and $\kappa_{ab}$ is not influenced by the phase transition, whereas an increase in $\kappa$ for polycrystalline BaFe$_2$As$_2$ is observed.  In the optimally doped composition, however, the excitations are two-dimensional and $\kappa_{ab}$ increases rapidly below the superconducting transition.  This detailed understanding of $\kappa(T)$ is made possible by the large amount of information already obtained through inelastic neutron scattering studies.  These results demonstrate, though, that the behavior of $\kappa(T)$ near $T_c$ may provide significant insight into the relative importance and/or nature of magnetic fluctuations.  As such, investigating $\kappa$ near $T_c$ is potentially useful in the screening of novel superconductors for unconventional pairing mechanisms.

\section{Acknowledgements}
We thank A. D. Christianson for useful discussions. This research was supported by the U.S. Department of Energy, Office of Science, Materials Sciences and Engineering Division.


\begin{thebibliography}{57}%
\makeatletter
\providecommand \@ifxundefined [1]{%
 \@ifx{#1\undefined}
}%
\providecommand \@ifnum [1]{%
 \ifnum #1\expandafter \@firstoftwo
 \else \expandafter \@secondoftwo
 \fi
}%
\providecommand \@ifx [1]{%
 \ifx #1\expandafter \@firstoftwo
 \else \expandafter \@secondoftwo
 \fi
}%
\providecommand \natexlab [1]{#1}%
\providecommand \enquote  [1]{``#1''}%
\providecommand \bibnamefont  [1]{#1}%
\providecommand \bibfnamefont [1]{#1}%
\providecommand \citenamefont [1]{#1}%
\providecommand \href@noop [0]{\@secondoftwo}%
\providecommand \href [0]{\begingroup \@sanitize@url \@href}%
\providecommand \@href[1]{\@@startlink{#1}\@@href}%
\providecommand \@@href[1]{\endgroup#1\@@endlink}%
\providecommand \@sanitize@url [0]{\catcode `\\12\catcode `\$12\catcode
  `\&12\catcode `\#12\catcode `\^12\catcode `\_12\catcode `\%12\relax}%
\providecommand \@@startlink[1]{}%
\providecommand \@@endlink[0]{}%
\providecommand \url  [0]{\begingroup\@sanitize@url \@url }%
\providecommand \@url [1]{\endgroup\@href {#1}{\urlprefix }}%
\providecommand \urlprefix  [0]{URL }%
\providecommand \Eprint [0]{\href }%
\providecommand \doibase [0]{http://dx.doi.org/}%
\providecommand \selectlanguage [0]{\@gobble}%
\providecommand \bibinfo  [0]{\@secondoftwo}%
\providecommand \bibfield  [0]{\@secondoftwo}%
\providecommand \translation [1]{[#1]}%
\providecommand \BibitemOpen [0]{}%
\providecommand \bibitemStop [0]{}%
\providecommand \bibitemNoStop [0]{.\EOS\space}%
\providecommand \EOS [0]{\spacefactor3000\relax}%
\providecommand \BibitemShut  [1]{\csname bibitem#1\endcsname}%
\let\auto@bib@innerbib\@empty
\bibitem [{\citenamefont {Ausloos}\ and\ \citenamefont
  {Houssa}(1999)}]{Ausloos1999}%
  \BibitemOpen
  \bibfield  {author} {\bibinfo {author} {\bibfnamefont {M.}~\bibnamefont
  {Ausloos}}\ and\ \bibinfo {author} {\bibfnamefont {M.}~\bibnamefont
  {Houssa}},\ }\href@noop {} {\bibfield  {journal} {\bibinfo  {journal}
  {Supercond. Sci. Technol.}\ }\textbf {\bibinfo {volume} {12}},\ \bibinfo
  {pages} {R103} (\bibinfo {year} {1999})}\BibitemShut {NoStop}%
\bibitem [{\citenamefont {Yu}\ \emph {et~al.}(1992)\citenamefont {Yu},
  \citenamefont {Salamon}, \citenamefont {Lu},\ and\ \citenamefont
  {Lee}}]{Yu1992}%
  \BibitemOpen
  \bibfield  {author} {\bibinfo {author} {\bibfnamefont {R.~C.}\ \bibnamefont
  {Yu}}, \bibinfo {author} {\bibfnamefont {M.~B.}\ \bibnamefont {Salamon}},
  \bibinfo {author} {\bibfnamefont {J.~P.}\ \bibnamefont {Lu}}, \ and\ \bibinfo
  {author} {\bibfnamefont {W.~C.}\ \bibnamefont {Lee}},\ }\href {\doibase
  10.1103/PhysRevLett.69.1431} {\bibfield  {journal} {\bibinfo  {journal}
  {Phys. Rev. Lett.}\ }\textbf {\bibinfo {volume} {69}},\ \bibinfo {pages}
  {1431} (\bibinfo {year} {1992})}\BibitemShut {NoStop}%
\bibitem [{\citenamefont {Pogorelov}\ \emph {et~al.}(1995)\citenamefont
  {Pogorelov}, \citenamefont {Arranz}, \citenamefont {Villar},\ and\
  \citenamefont {Vieira}}]{Pogorelov1995}%
  \BibitemOpen
  \bibfield  {author} {\bibinfo {author} {\bibfnamefont {Y.}~\bibnamefont
  {Pogorelov}}, \bibinfo {author} {\bibfnamefont {M.~A.}\ \bibnamefont
  {Arranz}}, \bibinfo {author} {\bibfnamefont {R.}~\bibnamefont {Villar}}, \
  and\ \bibinfo {author} {\bibfnamefont {S.}~\bibnamefont {Vieira}},\ }\href
  {\doibase 10.1103/PhysRevB.51.15474} {\bibfield  {journal} {\bibinfo
  {journal} {Phys. Rev. B}\ }\textbf {\bibinfo {volume} {51}},\ \bibinfo
  {pages} {15474} (\bibinfo {year} {1995})}\BibitemShut {NoStop}%
\bibitem [{\citenamefont {Zhang}\ \emph {et~al.}(2001)\citenamefont {Zhang},
  \citenamefont {Ong}, \citenamefont {Anderson}, \citenamefont {Bonn},
  \citenamefont {Liang},\ and\ \citenamefont {Hardy}}]{ZhangPRL2001}%
  \BibitemOpen
  \bibfield  {author} {\bibinfo {author} {\bibfnamefont {Y.}~\bibnamefont
  {Zhang}}, \bibinfo {author} {\bibfnamefont {N.~P.}\ \bibnamefont {Ong}},
  \bibinfo {author} {\bibfnamefont {P.~W.}\ \bibnamefont {Anderson}}, \bibinfo
  {author} {\bibfnamefont {D.~A.}\ \bibnamefont {Bonn}}, \bibinfo {author}
  {\bibfnamefont {R.}~\bibnamefont {Liang}}, \ and\ \bibinfo {author}
  {\bibfnamefont {W.~N.}\ \bibnamefont {Hardy}},\ }\href {\doibase
  10.1103/PhysRevLett.86.890} {\bibfield  {journal} {\bibinfo  {journal} {Phys.
  Rev. Lett.}\ }\textbf {\bibinfo {volume} {86}},\ \bibinfo {pages} {890}
  (\bibinfo {year} {2001})}\BibitemShut {NoStop}%
\bibitem [{\citenamefont {Movshovich}\ \emph {et~al.}(2001)\citenamefont
  {Movshovich}, \citenamefont {Jaime}, \citenamefont {Thompson}, \citenamefont
  {Petrovic}, \citenamefont {Fisk}, \citenamefont {Pagliuso},\ and\
  \citenamefont {Sarrao}}]{Movshovich2001}%
  \BibitemOpen
  \bibfield  {author} {\bibinfo {author} {\bibfnamefont {R.}~\bibnamefont
  {Movshovich}}, \bibinfo {author} {\bibfnamefont {M.}~\bibnamefont {Jaime}},
  \bibinfo {author} {\bibfnamefont {J.~D.}\ \bibnamefont {Thompson}}, \bibinfo
  {author} {\bibfnamefont {C.}~\bibnamefont {Petrovic}}, \bibinfo {author}
  {\bibfnamefont {Z.}~\bibnamefont {Fisk}}, \bibinfo {author} {\bibfnamefont
  {P.~G.}\ \bibnamefont {Pagliuso}}, \ and\ \bibinfo {author} {\bibfnamefont
  {J.~L.}\ \bibnamefont {Sarrao}},\ }\href {\doibase
  10.1103/PhysRevLett.86.5152} {\bibfield  {journal} {\bibinfo  {journal}
  {Phys. Rev. Lett.}\ }\textbf {\bibinfo {volume} {86}},\ \bibinfo {pages}
  {5152} (\bibinfo {year} {2001})}\BibitemShut {NoStop}%
\bibitem [{\citenamefont {Machida}\ \emph {et~al.}(2011)\citenamefont
  {Machida}, \citenamefont {Tomokuni}, \citenamefont {Isono}, \citenamefont
  {Izawa}, \citenamefont {Nakajima},\ and\ \citenamefont
  {Tamegai}}]{Machida2011}%
  \BibitemOpen
  \bibfield  {author} {\bibinfo {author} {\bibfnamefont {Y.}~\bibnamefont
  {Machida}}, \bibinfo {author} {\bibfnamefont {K.}~\bibnamefont {Tomokuni}},
  \bibinfo {author} {\bibfnamefont {T.}~\bibnamefont {Isono}}, \bibinfo
  {author} {\bibfnamefont {K.}~\bibnamefont {Izawa}}, \bibinfo {author}
  {\bibfnamefont {Y.}~\bibnamefont {Nakajima}}, \ and\ \bibinfo {author}
  {\bibfnamefont {T.}~\bibnamefont {Tamegai}},\ }\href {\doibase
  10.1016/j.physe.2010.07.036} {\bibfield  {journal} {\bibinfo  {journal}
  {Physica E}\ }\textbf {\bibinfo {volume} {43}},\ \bibinfo {pages} {714}
  (\bibinfo {year} {2011})}\BibitemShut {NoStop}%
\bibitem [{\citenamefont {Checkelsky}\ \emph {et~al.}(2012)\citenamefont
  {Checkelsky}, \citenamefont {Thomale}, \citenamefont {Li}, \citenamefont
  {Chen}, \citenamefont {Luo}, \citenamefont {Wang},\ and\ \citenamefont
  {Ong}}]{Checkelsky2012}%
  \BibitemOpen
  \bibfield  {author} {\bibinfo {author} {\bibfnamefont {J.~G.}\ \bibnamefont
  {Checkelsky}}, \bibinfo {author} {\bibfnamefont {R.}~\bibnamefont {Thomale}},
  \bibinfo {author} {\bibfnamefont {L.}~\bibnamefont {Li}}, \bibinfo {author}
  {\bibfnamefont {G.~F.}\ \bibnamefont {Chen}}, \bibinfo {author}
  {\bibfnamefont {J.~L.}\ \bibnamefont {Luo}}, \bibinfo {author} {\bibfnamefont
  {N.~L.}\ \bibnamefont {Wang}}, \ and\ \bibinfo {author} {\bibfnamefont
  {N.~P.}\ \bibnamefont {Ong}},\ }\href {\doibase 10.1103/PhysRevB.86.180502}
  {\bibfield  {journal} {\bibinfo  {journal} {Phys. Rev. B}\ }\textbf {\bibinfo
  {volume} {86}},\ \bibinfo {pages} {180502} (\bibinfo {year}
  {2012})}\BibitemShut {NoStop}%
\bibitem [{\citenamefont {Nakajima}\ \emph {et~al.}(2012)\citenamefont
  {Nakajima}, \citenamefont {Kurosaki},\ and\ \citenamefont
  {Tamegai}}]{Nakajima2012}%
  \BibitemOpen
  \bibfield  {author} {\bibinfo {author} {\bibfnamefont {Y.}~\bibnamefont
  {Nakajima}}, \bibinfo {author} {\bibfnamefont {Y.}~\bibnamefont {Kurosaki}},
  \ and\ \bibinfo {author} {\bibfnamefont {T.}~\bibnamefont {Tamegai}},\
  }\href@noop {} {\bibfield  {journal} {\bibinfo  {journal} {Journal of
  Physics: Conference Series}\ }\textbf {\bibinfo {volume} {400}},\ \bibinfo
  {pages} {022080} (\bibinfo {year} {2012})}\BibitemShut {NoStop}%
\bibitem [{\citenamefont {Bardeen}\ \emph {et~al.}(1959)\citenamefont
  {Bardeen}, \citenamefont {Rickayzen},\ and\ \citenamefont
  {Tewordt}}]{Bardeen1959}%
  \BibitemOpen
  \bibfield  {author} {\bibinfo {author} {\bibfnamefont {J.}~\bibnamefont
  {Bardeen}}, \bibinfo {author} {\bibfnamefont {G.}~\bibnamefont {Rickayzen}},
  \ and\ \bibinfo {author} {\bibfnamefont {L.}~\bibnamefont {Tewordt}},\ }\href
  {\doibase 10.1103/PhysRev.113.982} {\bibfield  {journal} {\bibinfo  {journal}
  {Phys. Rev.}\ }\textbf {\bibinfo {volume} {113}},\ \bibinfo {pages} {982}
  (\bibinfo {year} {1959})}\BibitemShut {NoStop}%
\bibitem [{\citenamefont {CODY}\ and\ \citenamefont {COHEN}(1964)}]{Cody1964}%
  \BibitemOpen
  \bibfield  {author} {\bibinfo {author} {\bibfnamefont {G.~D.}\ \bibnamefont
  {CODY}}\ and\ \bibinfo {author} {\bibfnamefont {R.~W.}\ \bibnamefont
  {COHEN}},\ }\href {\doibase 10.1103/RevModPhys.36.121} {\bibfield  {journal}
  {\bibinfo  {journal} {Rev. Mod. Phys.}\ }\textbf {\bibinfo {volume} {36}},\
  \bibinfo {pages} {121} (\bibinfo {year} {1964})}\BibitemShut {NoStop}%
\bibitem [{\citenamefont {Sefat}\ \emph {et~al.}(2008)\citenamefont {Sefat},
  \citenamefont {Jin}, \citenamefont {McGuire}, \citenamefont {Sales},
  \citenamefont {Singh},\ and\ \citenamefont {Mandrus}}]{Sefat2008}%
  \BibitemOpen
  \bibfield  {author} {\bibinfo {author} {\bibfnamefont {A.~S.}\ \bibnamefont
  {Sefat}}, \bibinfo {author} {\bibfnamefont {R.}~\bibnamefont {Jin}}, \bibinfo
  {author} {\bibfnamefont {M.~A.}\ \bibnamefont {McGuire}}, \bibinfo {author}
  {\bibfnamefont {B.~C.}\ \bibnamefont {Sales}}, \bibinfo {author}
  {\bibfnamefont {D.~J.}\ \bibnamefont {Singh}}, \ and\ \bibinfo {author}
  {\bibfnamefont {D.}~\bibnamefont {Mandrus}},\ }\href {\doibase
  10.1103/PhysRevLett.101.117004} {\bibfield  {journal} {\bibinfo  {journal}
  {Phys. Rev. Lett.}\ }\textbf {\bibinfo {volume} {101}},\ \bibinfo {pages}
  {117004} (\bibinfo {year} {2008})}\BibitemShut {NoStop}%
\bibitem [{\citenamefont {Rotter}\ \emph
  {et~al.}(2008{\natexlab{a}})\citenamefont {Rotter}, \citenamefont {Tegel},
  \citenamefont {Johrendt}, \citenamefont {Schellenberg}, \citenamefont
  {Hermes},\ and\ \citenamefont {P\"ottgen}}]{Rotter2008A}%
  \BibitemOpen
  \bibfield  {author} {\bibinfo {author} {\bibfnamefont {M.}~\bibnamefont
  {Rotter}}, \bibinfo {author} {\bibfnamefont {M.}~\bibnamefont {Tegel}},
  \bibinfo {author} {\bibfnamefont {D.}~\bibnamefont {Johrendt}}, \bibinfo
  {author} {\bibfnamefont {I.}~\bibnamefont {Schellenberg}}, \bibinfo {author}
  {\bibfnamefont {W.}~\bibnamefont {Hermes}}, \ and\ \bibinfo {author}
  {\bibfnamefont {R.}~\bibnamefont {P\"ottgen}},\ }\href {\doibase
  10.1103/PhysRevB.78.020503} {\bibfield  {journal} {\bibinfo  {journal} {Phys.
  Rev. B}\ }\textbf {\bibinfo {volume} {78}},\ \bibinfo {pages} {020503}
  (\bibinfo {year} {2008}{\natexlab{a}})}\BibitemShut {NoStop}%
\bibitem [{\citenamefont {Kim}\ \emph {et~al.}(2011)\citenamefont {Kim},
  \citenamefont {Fernandes}, \citenamefont {Kreyssig}, \citenamefont {Kim},
  \citenamefont {Thaler}, \citenamefont {Bud'ko}, \citenamefont {Canfield},
  \citenamefont {McQueeney}, \citenamefont {Schmalian},\ and\ \citenamefont
  {Goldman}}]{Kim2011b}%
  \BibitemOpen
  \bibfield  {author} {\bibinfo {author} {\bibfnamefont {M.~G.}\ \bibnamefont
  {Kim}}, \bibinfo {author} {\bibfnamefont {R.~M.}\ \bibnamefont {Fernandes}},
  \bibinfo {author} {\bibfnamefont {A.}~\bibnamefont {Kreyssig}}, \bibinfo
  {author} {\bibfnamefont {J.~W.}\ \bibnamefont {Kim}}, \bibinfo {author}
  {\bibfnamefont {A.}~\bibnamefont {Thaler}}, \bibinfo {author} {\bibfnamefont
  {S.~L.}\ \bibnamefont {Bud'ko}}, \bibinfo {author} {\bibfnamefont {P.~C.}\
  \bibnamefont {Canfield}}, \bibinfo {author} {\bibfnamefont {R.~J.}\
  \bibnamefont {McQueeney}}, \bibinfo {author} {\bibfnamefont {J.}~\bibnamefont
  {Schmalian}}, \ and\ \bibinfo {author} {\bibfnamefont {A.~I.}\ \bibnamefont
  {Goldman}},\ }\href {\doibase 10.1103/PhysRevB.83.134522} {\bibfield
  {journal} {\bibinfo  {journal} {Phys. Rev. B}\ }\textbf {\bibinfo {volume}
  {83}},\ \bibinfo {pages} {134522} (\bibinfo {year} {2011})}\BibitemShut
  {NoStop}%
\bibitem [{\citenamefont {Kitagawa}\ \emph {et~al.}(2008)\citenamefont
  {Kitagawa}, \citenamefont {Katayama}, \citenamefont {Ohgushi}, \citenamefont
  {Yoshida},\ and\ \citenamefont {Takigawa}}]{Kitagawa2008}%
  \BibitemOpen
  \bibfield  {author} {\bibinfo {author} {\bibfnamefont {K.}~\bibnamefont
  {Kitagawa}}, \bibinfo {author} {\bibfnamefont {N.}~\bibnamefont {Katayama}},
  \bibinfo {author} {\bibfnamefont {K.}~\bibnamefont {Ohgushi}}, \bibinfo
  {author} {\bibfnamefont {M.}~\bibnamefont {Yoshida}}, \ and\ \bibinfo
  {author} {\bibfnamefont {M.}~\bibnamefont {Takigawa}},\ }\href@noop {}
  {\bibfield  {journal} {\bibinfo  {journal} {J. Phys. Soc. Jpn.}\ }\textbf
  {\bibinfo {volume} {77}},\ \bibinfo {pages} {114709} (\bibinfo {year}
  {2008})}\BibitemShut {NoStop}%
\bibitem [{\citenamefont {Ning}\ \emph {et~al.}(2010)\citenamefont {Ning},
  \citenamefont {Ahilan}, \citenamefont {Imai}, \citenamefont {Sefat},
  \citenamefont {McGuire}, \citenamefont {Sales}, \citenamefont {Mandrus},
  \citenamefont {Cheng}, \citenamefont {Shen},\ and\ \citenamefont
  {Wen}}]{NingPRL2010}%
  \BibitemOpen
  \bibfield  {author} {\bibinfo {author} {\bibfnamefont {F.~L.}\ \bibnamefont
  {Ning}}, \bibinfo {author} {\bibfnamefont {K.}~\bibnamefont {Ahilan}},
  \bibinfo {author} {\bibfnamefont {T.}~\bibnamefont {Imai}}, \bibinfo {author}
  {\bibfnamefont {A.~S.}\ \bibnamefont {Sefat}}, \bibinfo {author}
  {\bibfnamefont {M.~A.}\ \bibnamefont {McGuire}}, \bibinfo {author}
  {\bibfnamefont {B.~C.}\ \bibnamefont {Sales}}, \bibinfo {author}
  {\bibfnamefont {D.}~\bibnamefont {Mandrus}}, \bibinfo {author} {\bibfnamefont
  {P.}~\bibnamefont {Cheng}}, \bibinfo {author} {\bibfnamefont
  {B.}~\bibnamefont {Shen}}, \ and\ \bibinfo {author} {\bibfnamefont {H.-H.}\
  \bibnamefont {Wen}},\ }\href {\doibase 10.1103/PhysRevLett.104.037001}
  {\bibfield  {journal} {\bibinfo  {journal} {Phys. Rev. Lett.}\ }\textbf
  {\bibinfo {volume} {104}},\ \bibinfo {pages} {037001} (\bibinfo {year}
  {2010})}\BibitemShut {NoStop}%
\bibitem [{\citenamefont {Pratt}\ \emph {et~al.}(2011)\citenamefont {Pratt},
  \citenamefont {Kim}, \citenamefont {Kreyssig}, \citenamefont {Lee},
  \citenamefont {Tucker}, \citenamefont {Thaler}, \citenamefont {Tian},
  \citenamefont {Zarestky}, \citenamefont {Bud'ko}, \citenamefont {Canfield},
  \citenamefont {Harmon}, \citenamefont {Goldman},\ and\ \citenamefont
  {McQueeney}}]{PrattPRL2011}%
  \BibitemOpen
  \bibfield  {author} {\bibinfo {author} {\bibfnamefont {D.~K.}\ \bibnamefont
  {Pratt}}, \bibinfo {author} {\bibfnamefont {M.~G.}\ \bibnamefont {Kim}},
  \bibinfo {author} {\bibfnamefont {A.}~\bibnamefont {Kreyssig}}, \bibinfo
  {author} {\bibfnamefont {Y.~B.}\ \bibnamefont {Lee}}, \bibinfo {author}
  {\bibfnamefont {G.~S.}\ \bibnamefont {Tucker}}, \bibinfo {author}
  {\bibfnamefont {A.}~\bibnamefont {Thaler}}, \bibinfo {author} {\bibfnamefont
  {W.}~\bibnamefont {Tian}}, \bibinfo {author} {\bibfnamefont {J.~L.}\
  \bibnamefont {Zarestky}}, \bibinfo {author} {\bibfnamefont {S.~L.}\
  \bibnamefont {Bud'ko}}, \bibinfo {author} {\bibfnamefont {P.~C.}\
  \bibnamefont {Canfield}}, \bibinfo {author} {\bibfnamefont {B.~N.}\
  \bibnamefont {Harmon}}, \bibinfo {author} {\bibfnamefont {A.~I.}\
  \bibnamefont {Goldman}}, \ and\ \bibinfo {author} {\bibfnamefont {R.~J.}\
  \bibnamefont {McQueeney}},\ }\href {\doibase 10.1103/PhysRevLett.106.257001}
  {\bibfield  {journal} {\bibinfo  {journal} {Phys. Rev. Lett.}\ }\textbf
  {\bibinfo {volume} {106}},\ \bibinfo {pages} {257001} (\bibinfo {year}
  {2011})}\BibitemShut {NoStop}%
\bibitem [{\citenamefont {Singh}(2009)}]{SinghReview2009}%
  \BibitemOpen
  \bibfield  {author} {\bibinfo {author} {\bibfnamefont {D.~J.}\ \bibnamefont
  {Singh}},\ }\href@noop {} {\bibfield  {journal} {\bibinfo  {journal} {Physica
  C: Supercond.}\ }\textbf {\bibinfo {volume} {469}},\ \bibinfo {pages} {418}
  (\bibinfo {year} {2009})}\BibitemShut {NoStop}%
\bibitem [{\citenamefont {Matan}\ \emph {et~al.}(2009)\citenamefont {Matan},
  \citenamefont {Morinaga}, \citenamefont {Iida},\ and\ \citenamefont
  {Sato}}]{MatanPRB2009}%
  \BibitemOpen
  \bibfield  {author} {\bibinfo {author} {\bibfnamefont {K.}~\bibnamefont
  {Matan}}, \bibinfo {author} {\bibfnamefont {R.}~\bibnamefont {Morinaga}},
  \bibinfo {author} {\bibfnamefont {K.}~\bibnamefont {Iida}}, \ and\ \bibinfo
  {author} {\bibfnamefont {T.~J.}\ \bibnamefont {Sato}},\ }\href {\doibase
  10.1103/PhysRevB.79.054526} {\bibfield  {journal} {\bibinfo  {journal} {Phys.
  Rev. B}\ }\textbf {\bibinfo {volume} {79}},\ \bibinfo {pages} {054526}
  (\bibinfo {year} {2009})}\BibitemShut {NoStop}%
\bibitem [{\citenamefont {Qureshi}\ \emph {et~al.}(2012)\citenamefont
  {Qureshi}, \citenamefont {Steffens}, \citenamefont {Wurmehl}, \citenamefont
  {Aswartham}, \citenamefont {B\"uchner},\ and\ \citenamefont
  {Braden}}]{Qureshi2012}%
  \BibitemOpen
  \bibfield  {author} {\bibinfo {author} {\bibfnamefont {N.}~\bibnamefont
  {Qureshi}}, \bibinfo {author} {\bibfnamefont {P.}~\bibnamefont {Steffens}},
  \bibinfo {author} {\bibfnamefont {S.}~\bibnamefont {Wurmehl}}, \bibinfo
  {author} {\bibfnamefont {S.}~\bibnamefont {Aswartham}}, \bibinfo {author}
  {\bibfnamefont {B.}~\bibnamefont {B\"uchner}}, \ and\ \bibinfo {author}
  {\bibfnamefont {M.}~\bibnamefont {Braden}},\ }\href {\doibase
  10.1103/PhysRevB.86.060410} {\bibfield  {journal} {\bibinfo  {journal} {Phys.
  Rev. B}\ }\textbf {\bibinfo {volume} {86}},\ \bibinfo {pages} {060410}
  (\bibinfo {year} {2012})}\BibitemShut {NoStop}%
\bibitem [{\citenamefont {Ni}\ \emph {et~al.}(2008)\citenamefont {Ni},
  \citenamefont {Tillman}, \citenamefont {Yan}, \citenamefont {Kracher},
  \citenamefont {Hannahs}, \citenamefont {Bud'ko},\ and\ \citenamefont
  {Canfield}}]{NiPRB2008}%
  \BibitemOpen
  \bibfield  {author} {\bibinfo {author} {\bibfnamefont {N.}~\bibnamefont
  {Ni}}, \bibinfo {author} {\bibfnamefont {M.~E.}\ \bibnamefont {Tillman}},
  \bibinfo {author} {\bibfnamefont {J.-Q.}\ \bibnamefont {Yan}}, \bibinfo
  {author} {\bibfnamefont {A.}~\bibnamefont {Kracher}}, \bibinfo {author}
  {\bibfnamefont {S.~T.}\ \bibnamefont {Hannahs}}, \bibinfo {author}
  {\bibfnamefont {S.~L.}\ \bibnamefont {Bud'ko}}, \ and\ \bibinfo {author}
  {\bibfnamefont {P.~C.}\ \bibnamefont {Canfield}},\ }\href {\doibase
  10.1103/PhysRevB.78.214515} {\bibfield  {journal} {\bibinfo  {journal} {Phys.
  Rev. B}\ }\textbf {\bibinfo {volume} {78}},\ \bibinfo {pages} {214515}
  (\bibinfo {year} {2008})}\BibitemShut {NoStop}%
\bibitem [{\citenamefont {Chu}\ \emph {et~al.}(2009)\citenamefont {Chu},
  \citenamefont {Analytis}, \citenamefont {Kucharczyk},\ and\ \citenamefont
  {Fisher}}]{ChuPRB2009}%
  \BibitemOpen
  \bibfield  {author} {\bibinfo {author} {\bibfnamefont {J.-H.}\ \bibnamefont
  {Chu}}, \bibinfo {author} {\bibfnamefont {J.~G.}\ \bibnamefont {Analytis}},
  \bibinfo {author} {\bibfnamefont {C.}~\bibnamefont {Kucharczyk}}, \ and\
  \bibinfo {author} {\bibfnamefont {I.~R.}\ \bibnamefont {Fisher}},\ }\href
  {\doibase 10.1103/PhysRevB.79.014506} {\bibfield  {journal} {\bibinfo
  {journal} {Phys. Rev. B}\ }\textbf {\bibinfo {volume} {79}},\ \bibinfo
  {pages} {014506} (\bibinfo {year} {2009})}\BibitemShut {NoStop}%
\bibitem [{\citenamefont {Lester}\ \emph {et~al.}(2009)\citenamefont {Lester},
  \citenamefont {Chu}, \citenamefont {Analytis}, \citenamefont {Capelli},
  \citenamefont {Erickson}, \citenamefont {Condron}, \citenamefont {Toney},
  \citenamefont {Fisher},\ and\ \citenamefont {Hayden}}]{Lester2009}%
  \BibitemOpen
  \bibfield  {author} {\bibinfo {author} {\bibfnamefont {C.}~\bibnamefont
  {Lester}}, \bibinfo {author} {\bibfnamefont {J.-H.}\ \bibnamefont {Chu}},
  \bibinfo {author} {\bibfnamefont {J.~G.}\ \bibnamefont {Analytis}}, \bibinfo
  {author} {\bibfnamefont {S.~C.}\ \bibnamefont {Capelli}}, \bibinfo {author}
  {\bibfnamefont {A.~S.}\ \bibnamefont {Erickson}}, \bibinfo {author}
  {\bibfnamefont {C.~L.}\ \bibnamefont {Condron}}, \bibinfo {author}
  {\bibfnamefont {M.~F.}\ \bibnamefont {Toney}}, \bibinfo {author}
  {\bibfnamefont {I.~R.}\ \bibnamefont {Fisher}}, \ and\ \bibinfo {author}
  {\bibfnamefont {S.~M.}\ \bibnamefont {Hayden}},\ }\href {\doibase
  10.1103/PhysRevB.79.144523} {\bibfield  {journal} {\bibinfo  {journal} {Phys.
  Rev. B}\ }\textbf {\bibinfo {volume} {79}},\ \bibinfo {pages} {144523}
  (\bibinfo {year} {2009})}\BibitemShut {NoStop}%
\bibitem [{\citenamefont {Canfield}\ \emph {et~al.}(2009)\citenamefont
  {Canfield}, \citenamefont {Bud'ko}, \citenamefont {Ni}, \citenamefont {Yan},\
  and\ \citenamefont {Kracher}}]{CanfieldPRB2009}%
  \BibitemOpen
  \bibfield  {author} {\bibinfo {author} {\bibfnamefont {P.~C.}\ \bibnamefont
  {Canfield}}, \bibinfo {author} {\bibfnamefont {S.~L.}\ \bibnamefont
  {Bud'ko}}, \bibinfo {author} {\bibfnamefont {N.}~\bibnamefont {Ni}}, \bibinfo
  {author} {\bibfnamefont {J.~Q.}\ \bibnamefont {Yan}}, \ and\ \bibinfo
  {author} {\bibfnamefont {A.}~\bibnamefont {Kracher}},\ }\href {\doibase
  10.1103/PhysRevB.80.060501} {\bibfield  {journal} {\bibinfo  {journal} {Phys.
  Rev. B}\ }\textbf {\bibinfo {volume} {80}},\ \bibinfo {pages} {060501}
  (\bibinfo {year} {2009})}\BibitemShut {NoStop}%
\bibitem [{\citenamefont {Mun}\ \emph {et~al.}(2009)\citenamefont {Mun},
  \citenamefont {Bud'ko}, \citenamefont {Ni}, \citenamefont {Thaler},\ and\
  \citenamefont {Canfield}}]{Mun2009}%
  \BibitemOpen
  \bibfield  {author} {\bibinfo {author} {\bibfnamefont {E.~D.}\ \bibnamefont
  {Mun}}, \bibinfo {author} {\bibfnamefont {S.~L.}\ \bibnamefont {Bud'ko}},
  \bibinfo {author} {\bibfnamefont {N.}~\bibnamefont {Ni}}, \bibinfo {author}
  {\bibfnamefont {A.~N.}\ \bibnamefont {Thaler}}, \ and\ \bibinfo {author}
  {\bibfnamefont {P.~C.}\ \bibnamefont {Canfield}},\ }\href {\doibase
  10.1103/PhysRevB.80.054517} {\bibfield  {journal} {\bibinfo  {journal} {Phys.
  Rev. B}\ }\textbf {\bibinfo {volume} {80}},\ \bibinfo {pages} {054517}
  (\bibinfo {year} {2009})}\BibitemShut {NoStop}%
\bibitem [{\citenamefont {Nandi}\ \emph {et~al.}(2010)\citenamefont {Nandi},
  \citenamefont {Kim}, \citenamefont {Kreyssig}, \citenamefont {Fernandes},
  \citenamefont {Pratt}, \citenamefont {Thaler}, \citenamefont {Ni},
  \citenamefont {Bud'ko}, \citenamefont {Canfield}, \citenamefont {Schmalian},
  \citenamefont {McQueeney},\ and\ \citenamefont {Goldman}}]{NandiPRL2010}%
  \BibitemOpen
  \bibfield  {author} {\bibinfo {author} {\bibfnamefont {S.}~\bibnamefont
  {Nandi}}, \bibinfo {author} {\bibfnamefont {M.~G.}\ \bibnamefont {Kim}},
  \bibinfo {author} {\bibfnamefont {A.}~\bibnamefont {Kreyssig}}, \bibinfo
  {author} {\bibfnamefont {R.~M.}\ \bibnamefont {Fernandes}}, \bibinfo {author}
  {\bibfnamefont {D.~K.}\ \bibnamefont {Pratt}}, \bibinfo {author}
  {\bibfnamefont {A.}~\bibnamefont {Thaler}}, \bibinfo {author} {\bibfnamefont
  {N.}~\bibnamefont {Ni}}, \bibinfo {author} {\bibfnamefont {S.~L.}\
  \bibnamefont {Bud'ko}}, \bibinfo {author} {\bibfnamefont {P.~C.}\
  \bibnamefont {Canfield}}, \bibinfo {author} {\bibfnamefont {J.}~\bibnamefont
  {Schmalian}}, \bibinfo {author} {\bibfnamefont {R.~J.}\ \bibnamefont
  {McQueeney}}, \ and\ \bibinfo {author} {\bibfnamefont {A.~I.}\ \bibnamefont
  {Goldman}},\ }\href {\doibase 10.1103/PhysRevLett.104.057006} {\bibfield
  {journal} {\bibinfo  {journal} {Phys. Rev. Lett.}\ }\textbf {\bibinfo
  {volume} {104}},\ \bibinfo {pages} {057006} (\bibinfo {year}
  {2010})}\BibitemShut {NoStop}%
\bibitem [{\citenamefont {Fernandes}\ \emph
  {et~al.}(2010{\natexlab{a}})\citenamefont {Fernandes}, \citenamefont {Pratt},
  \citenamefont {Tian}, \citenamefont {Zarestky}, \citenamefont {Kreyssig},
  \citenamefont {Nandi}, \citenamefont {Kim}, \citenamefont {Thaler},
  \citenamefont {Ni}, \citenamefont {Canfield}, \citenamefont {McQueeney},
  \citenamefont {Schmalian},\ and\ \citenamefont {Goldman}}]{FernandesPRB2010}%
  \BibitemOpen
  \bibfield  {author} {\bibinfo {author} {\bibfnamefont {R.~M.}\ \bibnamefont
  {Fernandes}}, \bibinfo {author} {\bibfnamefont {D.~K.}\ \bibnamefont
  {Pratt}}, \bibinfo {author} {\bibfnamefont {W.}~\bibnamefont {Tian}},
  \bibinfo {author} {\bibfnamefont {J.}~\bibnamefont {Zarestky}}, \bibinfo
  {author} {\bibfnamefont {A.}~\bibnamefont {Kreyssig}}, \bibinfo {author}
  {\bibfnamefont {S.}~\bibnamefont {Nandi}}, \bibinfo {author} {\bibfnamefont
  {M.~G.}\ \bibnamefont {Kim}}, \bibinfo {author} {\bibfnamefont
  {A.}~\bibnamefont {Thaler}}, \bibinfo {author} {\bibfnamefont
  {N.}~\bibnamefont {Ni}}, \bibinfo {author} {\bibfnamefont {P.~C.}\
  \bibnamefont {Canfield}}, \bibinfo {author} {\bibfnamefont {R.~J.}\
  \bibnamefont {McQueeney}}, \bibinfo {author} {\bibfnamefont {J.}~\bibnamefont
  {Schmalian}}, \ and\ \bibinfo {author} {\bibfnamefont {A.~I.}\ \bibnamefont
  {Goldman}},\ }\href {\doibase 10.1103/PhysRevB.81.140501} {\bibfield
  {journal} {\bibinfo  {journal} {Phys. Rev. B}\ }\textbf {\bibinfo {volume}
  {81}},\ \bibinfo {pages} {140501} (\bibinfo {year}
  {2010}{\natexlab{a}})}\BibitemShut {NoStop}%
\bibitem [{\citenamefont {Mandrus}\ \emph {et~al.}(2010)\citenamefont
  {Mandrus}, \citenamefont {Sefat}, \citenamefont {McGuire},\ and\
  \citenamefont {Sales}}]{Mandrus2010}%
  \BibitemOpen
  \bibfield  {author} {\bibinfo {author} {\bibfnamefont {D.}~\bibnamefont
  {Mandrus}}, \bibinfo {author} {\bibfnamefont {A.~S.}\ \bibnamefont {Sefat}},
  \bibinfo {author} {\bibfnamefont {M.~A.}\ \bibnamefont {McGuire}}, \ and\
  \bibinfo {author} {\bibfnamefont {B.~C.}\ \bibnamefont {Sales}},\ }\href@noop
  {} {\bibfield  {journal} {\bibinfo  {journal} {Chem. Mater.}\ }\textbf
  {\bibinfo {volume} {22}},\ \bibinfo {pages} {715} (\bibinfo {year}
  {2010})}\BibitemShut {NoStop}%
\bibitem [{\citenamefont {Johnston}(2010)}]{JohnstonReview2010}%
  \BibitemOpen
  \bibfield  {author} {\bibinfo {author} {\bibfnamefont {D.~C.}\ \bibnamefont
  {Johnston}},\ }\href@noop {} {\bibfield  {journal} {\bibinfo  {journal}
  {Advances in Physics}\ }\textbf {\bibinfo {volume} {59}},\ \bibinfo {pages}
  {803} (\bibinfo {year} {2010})}\BibitemShut {NoStop}%
\bibitem [{\citenamefont {Paglione}\ and\ \citenamefont
  {Greene}(2010)}]{PaglioneGreen2010}%
  \BibitemOpen
  \bibfield  {author} {\bibinfo {author} {\bibfnamefont {J.}~\bibnamefont
  {Paglione}}\ and\ \bibinfo {author} {\bibfnamefont {R.~L.}\ \bibnamefont
  {Greene}},\ }\href@noop {} {\bibfield  {journal} {\bibinfo  {journal} {Nature
  Physics}\ }\textbf {\bibinfo {volume} {6}},\ \bibinfo {pages} {645} (\bibinfo
  {year} {2010})}\BibitemShut {NoStop}%
\bibitem [{\citenamefont {Lumsden}\ and\ \citenamefont
  {Christianson}(2010)}]{LumsdenChristianson2010}%
  \BibitemOpen
  \bibfield  {author} {\bibinfo {author} {\bibfnamefont {M.~D.}\ \bibnamefont
  {Lumsden}}\ and\ \bibinfo {author} {\bibfnamefont {A.~D.}\ \bibnamefont
  {Christianson}},\ }\href@noop {} {\bibfield  {journal} {\bibinfo  {journal}
  {J. Phys.: Condens. Matter}\ }\textbf {\bibinfo {volume} {22}},\ \bibinfo
  {pages} {203203} (\bibinfo {year} {2010})}\BibitemShut {NoStop}%
\bibitem [{\citenamefont {Rullier-Albenque}\ \emph {et~al.}(2009)\citenamefont
  {Rullier-Albenque}, \citenamefont {Colson}, \citenamefont {Forget},\ and\
  \citenamefont {Alloul}}]{Albenque2009}%
  \BibitemOpen
  \bibfield  {author} {\bibinfo {author} {\bibfnamefont {F.}~\bibnamefont
  {Rullier-Albenque}}, \bibinfo {author} {\bibfnamefont {D.}~\bibnamefont
  {Colson}}, \bibinfo {author} {\bibfnamefont {A.}~\bibnamefont {Forget}}, \
  and\ \bibinfo {author} {\bibfnamefont {H.}~\bibnamefont {Alloul}},\ }\href
  {\doibase 10.1103/PhysRevLett.103.057001} {\bibfield  {journal} {\bibinfo
  {journal} {Phys. Rev. Lett.}\ }\textbf {\bibinfo {volume} {103}},\ \bibinfo
  {pages} {057001} (\bibinfo {year} {2009})}\BibitemShut {NoStop}%
\bibitem [{\citenamefont {Rotter}\ \emph
  {et~al.}(2008{\natexlab{b}})\citenamefont {Rotter}, \citenamefont {Tegel},\
  and\ \citenamefont {Johrendt}}]{Rotter2008B}%
  \BibitemOpen
  \bibfield  {author} {\bibinfo {author} {\bibfnamefont {M.}~\bibnamefont
  {Rotter}}, \bibinfo {author} {\bibfnamefont {M.}~\bibnamefont {Tegel}}, \
  and\ \bibinfo {author} {\bibfnamefont {D.}~\bibnamefont {Johrendt}},\ }\href
  {\doibase 10.1103/PhysRevLett.101.107006} {\bibfield  {journal} {\bibinfo
  {journal} {Phys. Rev. Lett.}\ }\textbf {\bibinfo {volume} {101}},\ \bibinfo
  {pages} {107006} (\bibinfo {year} {2008}{\natexlab{b}})}\BibitemShut
  {NoStop}%
\bibitem [{\citenamefont {Chen}\ \emph {et~al.}(2009)\citenamefont {Chen},
  \citenamefont {Ren}, \citenamefont {Qiu}, \citenamefont {Bao}, \citenamefont
  {Liu}, \citenamefont {Wu}, \citenamefont {Wu}, \citenamefont {Xie},
  \citenamefont {Huang},\ and\ \citenamefont {Chen}}]{ChenEPL2009}%
  \BibitemOpen
  \bibfield  {author} {\bibinfo {author} {\bibfnamefont {H.}~\bibnamefont
  {Chen}}, \bibinfo {author} {\bibfnamefont {Y.}~\bibnamefont {Ren}}, \bibinfo
  {author} {\bibfnamefont {Y.}~\bibnamefont {Qiu}}, \bibinfo {author}
  {\bibfnamefont {W.}~\bibnamefont {Bao}}, \bibinfo {author} {\bibfnamefont
  {R.~H.}\ \bibnamefont {Liu}}, \bibinfo {author} {\bibfnamefont
  {G.}~\bibnamefont {Wu}}, \bibinfo {author} {\bibfnamefont {T.}~\bibnamefont
  {Wu}}, \bibinfo {author} {\bibfnamefont {Y.~L.}\ \bibnamefont {Xie}},
  \bibinfo {author} {\bibfnamefont {X.~F. W.~Q.}\ \bibnamefont {Huang}}, \ and\
  \bibinfo {author} {\bibfnamefont {X.~H.}\ \bibnamefont {Chen}},\ }\href@noop
  {} {\bibfield  {journal} {\bibinfo  {journal} {Euro. Phys. Lett.}\ }\textbf
  {\bibinfo {volume} {85}},\ \bibinfo {pages} {17006} (\bibinfo {year}
  {2009})}\BibitemShut {NoStop}%
\bibitem [{\citenamefont {Jiang}\ \emph {et~al.}(2009)\citenamefont {Jiang},
  \citenamefont {Xing}, \citenamefont {Xuan}, \citenamefont {Wang},
  \citenamefont {Ren}, \citenamefont {Feng}, \citenamefont {Dai}, \citenamefont
  {Xu},\ and\ \citenamefont {Cao}}]{Jiang2009}%
  \BibitemOpen
  \bibfield  {author} {\bibinfo {author} {\bibfnamefont {S.}~\bibnamefont
  {Jiang}}, \bibinfo {author} {\bibfnamefont {H.}~\bibnamefont {Xing}},
  \bibinfo {author} {\bibfnamefont {G.}~\bibnamefont {Xuan}}, \bibinfo {author}
  {\bibfnamefont {C.}~\bibnamefont {Wang}}, \bibinfo {author} {\bibfnamefont
  {Z.}~\bibnamefont {Ren}}, \bibinfo {author} {\bibfnamefont {C.}~\bibnamefont
  {Feng}}, \bibinfo {author} {\bibfnamefont {J.}~\bibnamefont {Dai}}, \bibinfo
  {author} {\bibfnamefont {Z.}~\bibnamefont {Xu}}, \ and\ \bibinfo {author}
  {\bibfnamefont {G.}~\bibnamefont {Cao}},\ }\href@noop {} {\bibfield
  {journal} {\bibinfo  {journal} {J. Phys.: Condens. Matter}\ }\textbf
  {\bibinfo {volume} {21}},\ \bibinfo {pages} {382203} (\bibinfo {year}
  {2009})}\BibitemShut {NoStop}%
\bibitem [{\citenamefont {Christianson}\ \emph {et~al.}(2009)\citenamefont
  {Christianson}, \citenamefont {Lumsden}, \citenamefont {Nagler},
  \citenamefont {MacDougall}, \citenamefont {McGuire}, \citenamefont {Sefat},
  \citenamefont {Jin}, \citenamefont {Sales},\ and\ \citenamefont
  {Mandrus}}]{ChristiansonPRL2009}%
  \BibitemOpen
  \bibfield  {author} {\bibinfo {author} {\bibfnamefont {A.~D.}\ \bibnamefont
  {Christianson}}, \bibinfo {author} {\bibfnamefont {M.~D.}\ \bibnamefont
  {Lumsden}}, \bibinfo {author} {\bibfnamefont {S.~E.}\ \bibnamefont {Nagler}},
  \bibinfo {author} {\bibfnamefont {G.~J.}\ \bibnamefont {MacDougall}},
  \bibinfo {author} {\bibfnamefont {M.~A.}\ \bibnamefont {McGuire}}, \bibinfo
  {author} {\bibfnamefont {A.~S.}\ \bibnamefont {Sefat}}, \bibinfo {author}
  {\bibfnamefont {R.}~\bibnamefont {Jin}}, \bibinfo {author} {\bibfnamefont
  {B.~C.}\ \bibnamefont {Sales}}, \ and\ \bibinfo {author} {\bibfnamefont
  {D.}~\bibnamefont {Mandrus}},\ }\href {\doibase
  10.1103/PhysRevLett.103.087002} {\bibfield  {journal} {\bibinfo  {journal}
  {Phys. Rev. Lett.}\ }\textbf {\bibinfo {volume} {103}},\ \bibinfo {pages}
  {087002} (\bibinfo {year} {2009})}\BibitemShut {NoStop}%
\bibitem [{\citenamefont {Pratt}\ \emph {et~al.}(2009)\citenamefont {Pratt},
  \citenamefont {Tian}, \citenamefont {Kreyssig}, \citenamefont {Zarestky},
  \citenamefont {Nandi}, \citenamefont {Ni}, \citenamefont {Bud'ko},
  \citenamefont {Canfield}, \citenamefont {Goldman},\ and\ \citenamefont
  {McQueeney}}]{PrattPRL2009}%
  \BibitemOpen
  \bibfield  {author} {\bibinfo {author} {\bibfnamefont {D.~K.}\ \bibnamefont
  {Pratt}}, \bibinfo {author} {\bibfnamefont {W.}~\bibnamefont {Tian}},
  \bibinfo {author} {\bibfnamefont {A.}~\bibnamefont {Kreyssig}}, \bibinfo
  {author} {\bibfnamefont {J.~L.}\ \bibnamefont {Zarestky}}, \bibinfo {author}
  {\bibfnamefont {S.}~\bibnamefont {Nandi}}, \bibinfo {author} {\bibfnamefont
  {N.}~\bibnamefont {Ni}}, \bibinfo {author} {\bibfnamefont {S.~L.}\
  \bibnamefont {Bud'ko}}, \bibinfo {author} {\bibfnamefont {P.~C.}\
  \bibnamefont {Canfield}}, \bibinfo {author} {\bibfnamefont {A.~I.}\
  \bibnamefont {Goldman}}, \ and\ \bibinfo {author} {\bibfnamefont {R.~J.}\
  \bibnamefont {McQueeney}},\ }\href {\doibase 10.1103/PhysRevLett.103.087001}
  {\bibfield  {journal} {\bibinfo  {journal} {Phys. Rev. Lett.}\ }\textbf
  {\bibinfo {volume} {103}},\ \bibinfo {pages} {087001} (\bibinfo {year}
  {2009})}\BibitemShut {NoStop}%
\bibitem [{\citenamefont {Mazin}\ \emph {et~al.}(2008)\citenamefont {Mazin},
  \citenamefont {Johannes}, \citenamefont {Boeri}, \citenamefont {Koepernik},\
  and\ \citenamefont {Singh}}]{MazinPRB2008}%
  \BibitemOpen
  \bibfield  {author} {\bibinfo {author} {\bibfnamefont {I.~I.}\ \bibnamefont
  {Mazin}}, \bibinfo {author} {\bibfnamefont {M.~D.}\ \bibnamefont {Johannes}},
  \bibinfo {author} {\bibfnamefont {L.}~\bibnamefont {Boeri}}, \bibinfo
  {author} {\bibfnamefont {K.}~\bibnamefont {Koepernik}}, \ and\ \bibinfo
  {author} {\bibfnamefont {D.~J.}\ \bibnamefont {Singh}},\ }\href {\doibase
  10.1103/PhysRevB.78.085104} {\bibfield  {journal} {\bibinfo  {journal} {Phys.
  Rev. B}\ }\textbf {\bibinfo {volume} {78}},\ \bibinfo {pages} {085104}
  (\bibinfo {year} {2008})}\BibitemShut {NoStop}%
\bibitem [{\citenamefont {Lumsden}\ \emph {et~al.}(2009)\citenamefont
  {Lumsden}, \citenamefont {Christianson}, \citenamefont {Parshall},
  \citenamefont {Stone}, \citenamefont {Nagler}, \citenamefont {MacDougall},
  \citenamefont {Mook}, \citenamefont {Lokshin}, \citenamefont {Egami},
  \citenamefont {Abernathy}, \citenamefont {Goremychkin}, \citenamefont
  {Osborn}, \citenamefont {McGuire}, \citenamefont {Sefat}, \citenamefont
  {Jin}, \citenamefont {Sales},\ and\ \citenamefont
  {Mandrus}}]{LumsdenPRL2009}%
  \BibitemOpen
  \bibfield  {author} {\bibinfo {author} {\bibfnamefont {M.~D.}\ \bibnamefont
  {Lumsden}}, \bibinfo {author} {\bibfnamefont {A.~D.}\ \bibnamefont
  {Christianson}}, \bibinfo {author} {\bibfnamefont {D.}~\bibnamefont
  {Parshall}}, \bibinfo {author} {\bibfnamefont {M.~B.}\ \bibnamefont {Stone}},
  \bibinfo {author} {\bibfnamefont {S.~E.}\ \bibnamefont {Nagler}}, \bibinfo
  {author} {\bibfnamefont {G.~J.}\ \bibnamefont {MacDougall}}, \bibinfo
  {author} {\bibfnamefont {H.~A.}\ \bibnamefont {Mook}}, \bibinfo {author}
  {\bibfnamefont {K.}~\bibnamefont {Lokshin}}, \bibinfo {author} {\bibfnamefont
  {T.}~\bibnamefont {Egami}}, \bibinfo {author} {\bibfnamefont {D.~L.}\
  \bibnamefont {Abernathy}}, \bibinfo {author} {\bibfnamefont {E.~A.}\
  \bibnamefont {Goremychkin}}, \bibinfo {author} {\bibfnamefont
  {R.}~\bibnamefont {Osborn}}, \bibinfo {author} {\bibfnamefont {M.~A.}\
  \bibnamefont {McGuire}}, \bibinfo {author} {\bibfnamefont {A.~S.}\
  \bibnamefont {Sefat}}, \bibinfo {author} {\bibfnamefont {R.}~\bibnamefont
  {Jin}}, \bibinfo {author} {\bibfnamefont {B.~C.}\ \bibnamefont {Sales}}, \
  and\ \bibinfo {author} {\bibfnamefont {D.}~\bibnamefont {Mandrus}},\ }\href
  {\doibase 10.1103/PhysRevLett.102.107005} {\bibfield  {journal} {\bibinfo
  {journal} {Phys. Rev. Lett.}\ }\textbf {\bibinfo {volume} {102}},\ \bibinfo
  {pages} {107005} (\bibinfo {year} {2009})}\BibitemShut {NoStop}%
\bibitem [{\citenamefont {Inosov}\ \emph {et~al.}(2010)\citenamefont {Inosov},
  \citenamefont {Park}, \citenamefont {Bourges}, \citenamefont {Sun},
  \citenamefont {Sidis}, \citenamefont {Schneidewind}, \citenamefont {Hradil},
  \citenamefont {Haug}, \citenamefont {Lin}, \citenamefont {Keimer},\ and\
  \citenamefont {Hinkov}}]{Inosov2009}%
  \BibitemOpen
  \bibfield  {author} {\bibinfo {author} {\bibfnamefont {D.~S.}\ \bibnamefont
  {Inosov}}, \bibinfo {author} {\bibfnamefont {J.~T.}\ \bibnamefont {Park}},
  \bibinfo {author} {\bibfnamefont {P.}~\bibnamefont {Bourges}}, \bibinfo
  {author} {\bibfnamefont {D.~L.}\ \bibnamefont {Sun}}, \bibinfo {author}
  {\bibfnamefont {Y.}~\bibnamefont {Sidis}}, \bibinfo {author} {\bibfnamefont
  {A.}~\bibnamefont {Schneidewind}}, \bibinfo {author} {\bibfnamefont
  {K.}~\bibnamefont {Hradil}}, \bibinfo {author} {\bibfnamefont
  {D.}~\bibnamefont {Haug}}, \bibinfo {author} {\bibfnamefont {C.~T.}\
  \bibnamefont {Lin}}, \bibinfo {author} {\bibfnamefont {B.}~\bibnamefont
  {Keimer}}, \ and\ \bibinfo {author} {\bibfnamefont {V.}~\bibnamefont
  {Hinkov}},\ }\href@noop {} {\bibfield  {journal} {\bibinfo  {journal} {Nature
  Physics}\ }\textbf {\bibinfo {volume} {6}},\ \bibinfo {pages} {178} (\bibinfo
  {year} {2010})}\BibitemShut {NoStop}%
\bibitem [{\citenamefont {H{\"u}fner}\ \emph {et~al.}(2008)\citenamefont
  {H{\"u}fner}, \citenamefont {Hossain}, \citenamefont {Damascelli},\ and\
  \citenamefont {Sawatzky}}]{Hufner2008}%
  \BibitemOpen
  \bibfield  {author} {\bibinfo {author} {\bibfnamefont {S.}~\bibnamefont
  {H{\"u}fner}}, \bibinfo {author} {\bibfnamefont {M.~A.}\ \bibnamefont
  {Hossain}}, \bibinfo {author} {\bibfnamefont {A.}~\bibnamefont {Damascelli}},
  \ and\ \bibinfo {author} {\bibfnamefont {G.~A.}\ \bibnamefont {Sawatzky}},\
  }\href@noop {} {\bibfield  {journal} {\bibinfo  {journal} {Rep. Prog. Phys.}\
  }\textbf {\bibinfo {volume} {71}},\ \bibinfo {pages} {062501} (\bibinfo
  {year} {2008})}\BibitemShut {NoStop}%
\bibitem [{\citenamefont {Lester}\ \emph {et~al.}(2010)\citenamefont {Lester},
  \citenamefont {Chu}, \citenamefont {Analytis}, \citenamefont {Perring},
  \citenamefont {Fisher},\ and\ \citenamefont {Hayden}}]{Lester2010}%
  \BibitemOpen
  \bibfield  {author} {\bibinfo {author} {\bibfnamefont {C.}~\bibnamefont
  {Lester}}, \bibinfo {author} {\bibfnamefont {J.-H.}\ \bibnamefont {Chu}},
  \bibinfo {author} {\bibfnamefont {J.~G.}\ \bibnamefont {Analytis}}, \bibinfo
  {author} {\bibfnamefont {T.~G.}\ \bibnamefont {Perring}}, \bibinfo {author}
  {\bibfnamefont {I.~R.}\ \bibnamefont {Fisher}}, \ and\ \bibinfo {author}
  {\bibfnamefont {S.~M.}\ \bibnamefont {Hayden}},\ }\href {\doibase
  10.1103/PhysRevB.81.064505} {\bibfield  {journal} {\bibinfo  {journal} {Phys.
  Rev. B}\ }\textbf {\bibinfo {volume} {81}},\ \bibinfo {pages} {064505}
  (\bibinfo {year} {2010})}\BibitemShut {NoStop}%
\bibitem [{\citenamefont {Tucker}\ \emph {et~al.}(2012)\citenamefont {Tucker},
  \citenamefont {Fernandes}, \citenamefont {Li}, \citenamefont {Thampy},
  \citenamefont {Ni}, \citenamefont {Abernathy}, \citenamefont {Bud'ko},
  \citenamefont {Canfield}, \citenamefont {Vaknin}, \citenamefont {Schmalian},\
  and\ \citenamefont {McQueeney}}]{TuckerPRB2012}%
  \BibitemOpen
  \bibfield  {author} {\bibinfo {author} {\bibfnamefont {G.~S.}\ \bibnamefont
  {Tucker}}, \bibinfo {author} {\bibfnamefont {R.~M.}\ \bibnamefont
  {Fernandes}}, \bibinfo {author} {\bibfnamefont {H.-F.}\ \bibnamefont {Li}},
  \bibinfo {author} {\bibfnamefont {V.}~\bibnamefont {Thampy}}, \bibinfo
  {author} {\bibfnamefont {N.}~\bibnamefont {Ni}}, \bibinfo {author}
  {\bibfnamefont {D.~L.}\ \bibnamefont {Abernathy}}, \bibinfo {author}
  {\bibfnamefont {S.~L.}\ \bibnamefont {Bud'ko}}, \bibinfo {author}
  {\bibfnamefont {P.~C.}\ \bibnamefont {Canfield}}, \bibinfo {author}
  {\bibfnamefont {D.}~\bibnamefont {Vaknin}}, \bibinfo {author} {\bibfnamefont
  {J.}~\bibnamefont {Schmalian}}, \ and\ \bibinfo {author} {\bibfnamefont
  {R.~J.}\ \bibnamefont {McQueeney}},\ }\href {\doibase
  10.1103/PhysRevB.86.024505} {\bibfield  {journal} {\bibinfo  {journal} {Phys.
  Rev. B}\ }\textbf {\bibinfo {volume} {86}},\ \bibinfo {pages} {024505}
  (\bibinfo {year} {2012})}\BibitemShut {NoStop}%
\bibitem [{\citenamefont {Matan}\ \emph {et~al.}(2010)\citenamefont {Matan},
  \citenamefont {Ibuka}, \citenamefont {Morinaga}, \citenamefont {Chi},
  \citenamefont {Lynn}, \citenamefont {Christianson}, \citenamefont {Lumsden},\
  and\ \citenamefont {Sato}}]{MatanPRB2010}%
  \BibitemOpen
  \bibfield  {author} {\bibinfo {author} {\bibfnamefont {K.}~\bibnamefont
  {Matan}}, \bibinfo {author} {\bibfnamefont {S.}~\bibnamefont {Ibuka}},
  \bibinfo {author} {\bibfnamefont {R.}~\bibnamefont {Morinaga}}, \bibinfo
  {author} {\bibfnamefont {S.}~\bibnamefont {Chi}}, \bibinfo {author}
  {\bibfnamefont {J.~W.}\ \bibnamefont {Lynn}}, \bibinfo {author}
  {\bibfnamefont {A.~D.}\ \bibnamefont {Christianson}}, \bibinfo {author}
  {\bibfnamefont {M.~D.}\ \bibnamefont {Lumsden}}, \ and\ \bibinfo {author}
  {\bibfnamefont {T.~J.}\ \bibnamefont {Sato}},\ }\href {\doibase
  10.1103/PhysRevB.82.054515} {\bibfield  {journal} {\bibinfo  {journal} {Phys.
  Rev. B}\ }\textbf {\bibinfo {volume} {82}},\ \bibinfo {pages} {054515}
  (\bibinfo {year} {2010})}\BibitemShut {NoStop}%
\bibitem [{\citenamefont {Gofryk}\ \emph {et~al.}(2011)\citenamefont {Gofryk},
  \citenamefont {Vorontsov}, \citenamefont {Vekhter}, \citenamefont {Sefat},
  \citenamefont {Imai}, \citenamefont {Bauer}, \citenamefont {Thompson},\ and\
  \citenamefont {Ronning}}]{Gofryk2011}%
  \BibitemOpen
  \bibfield  {author} {\bibinfo {author} {\bibfnamefont {K.}~\bibnamefont
  {Gofryk}}, \bibinfo {author} {\bibfnamefont {A.~B.}\ \bibnamefont
  {Vorontsov}}, \bibinfo {author} {\bibfnamefont {I.}~\bibnamefont {Vekhter}},
  \bibinfo {author} {\bibfnamefont {A.~S.}\ \bibnamefont {Sefat}}, \bibinfo
  {author} {\bibfnamefont {T.}~\bibnamefont {Imai}}, \bibinfo {author}
  {\bibfnamefont {E.~D.}\ \bibnamefont {Bauer}}, \bibinfo {author}
  {\bibfnamefont {J.~D.}\ \bibnamefont {Thompson}}, \ and\ \bibinfo {author}
  {\bibfnamefont {F.}~\bibnamefont {Ronning}},\ }\href {\doibase
  10.1103/PhysRevB.83.064513} {\bibfield  {journal} {\bibinfo  {journal} {Phys.
  Rev. B}\ }\textbf {\bibinfo {volume} {83}},\ \bibinfo {pages} {064513}
  (\bibinfo {year} {2011})}\BibitemShut {NoStop}%
\bibitem [{\citenamefont {Sharma}\ \emph {et~al.}(2004)\citenamefont {Sharma},
  \citenamefont {Ahn}, \citenamefont {Hur}, \citenamefont {Park}, \citenamefont
  {Kim}, \citenamefont {Lee}, \citenamefont {Park}, \citenamefont {Guha},\ and\
  \citenamefont {Cheong}}]{SharmaPRL2004}%
  \BibitemOpen
  \bibfield  {author} {\bibinfo {author} {\bibfnamefont {P.~A.}\ \bibnamefont
  {Sharma}}, \bibinfo {author} {\bibfnamefont {J.~S.}\ \bibnamefont {Ahn}},
  \bibinfo {author} {\bibfnamefont {N.}~\bibnamefont {Hur}}, \bibinfo {author}
  {\bibfnamefont {S.}~\bibnamefont {Park}}, \bibinfo {author} {\bibfnamefont
  {S.~B.}\ \bibnamefont {Kim}}, \bibinfo {author} {\bibfnamefont
  {S.}~\bibnamefont {Lee}}, \bibinfo {author} {\bibfnamefont {J.-G.}\
  \bibnamefont {Park}}, \bibinfo {author} {\bibfnamefont {S.}~\bibnamefont
  {Guha}}, \ and\ \bibinfo {author} {\bibfnamefont {S.-W.}\ \bibnamefont
  {Cheong}},\ }\href {\doibase 10.1103/PhysRevLett.93.177202} {\bibfield
  {journal} {\bibinfo  {journal} {Phys. Rev. Lett.}\ }\textbf {\bibinfo
  {volume} {93}},\ \bibinfo {pages} {177202} (\bibinfo {year}
  {2004})}\BibitemShut {NoStop}%
\bibitem [{\citenamefont {Bhandari}\ and\ \citenamefont
  {Verma}(1966)}]{Bhandari1966}%
  \BibitemOpen
  \bibfield  {author} {\bibinfo {author} {\bibfnamefont {C.~M.}\ \bibnamefont
  {Bhandari}}\ and\ \bibinfo {author} {\bibfnamefont {G.~S.}\ \bibnamefont
  {Verma}},\ }\href {\doibase 10.1103/PhysRev.152.731} {\bibfield  {journal}
  {\bibinfo  {journal} {Phys. Rev.}\ }\textbf {\bibinfo {volume} {152}},\
  \bibinfo {pages} {731} (\bibinfo {year} {1966})}\BibitemShut {NoStop}%
\bibitem [{\citenamefont {Sales}\ \emph {et~al.}(2012)\citenamefont {Sales},
  \citenamefont {May}, \citenamefont {McGuire}, \citenamefont {Stone},
  \citenamefont {Singh},\ and\ \citenamefont {Mandrus}}]{SalesCrSb2012}%
  \BibitemOpen
  \bibfield  {author} {\bibinfo {author} {\bibfnamefont {B.~C.}\ \bibnamefont
  {Sales}}, \bibinfo {author} {\bibfnamefont {A.~F.}\ \bibnamefont {May}},
  \bibinfo {author} {\bibfnamefont {M.~A.}\ \bibnamefont {McGuire}}, \bibinfo
  {author} {\bibfnamefont {M.~B.}\ \bibnamefont {Stone}}, \bibinfo {author}
  {\bibfnamefont {D.~J.}\ \bibnamefont {Singh}}, \ and\ \bibinfo {author}
  {\bibfnamefont {D.}~\bibnamefont {Mandrus}},\ }\href {\doibase
  10.1103/PhysRevB.86.235136} {\bibfield  {journal} {\bibinfo  {journal} {Phys.
  Rev. B}\ }\textbf {\bibinfo {volume} {86}},\ \bibinfo {pages} {235136}
  (\bibinfo {year} {2012})}\BibitemShut {NoStop}%
\bibitem [{\citenamefont {McGuire}\ \emph {et~al.}(2008)\citenamefont
  {McGuire}, \citenamefont {Christianson}, \citenamefont {Sefat}, \citenamefont
  {Sales}, \citenamefont {Lumsden}, \citenamefont {Jin}, \citenamefont
  {Payzant}, \citenamefont {Mandrus}, \citenamefont {Luan}, \citenamefont
  {Keppens}, \citenamefont {Varadarajan}, \citenamefont {Brill}, \citenamefont
  {Hermann}, \citenamefont {Sougrati}, \citenamefont {Grandjean},\ and\
  \citenamefont {Long}}]{McGuire_LaFeAsO}%
  \BibitemOpen
  \bibfield  {author} {\bibinfo {author} {\bibfnamefont {M.~A.}\ \bibnamefont
  {McGuire}}, \bibinfo {author} {\bibfnamefont {A.~D.}\ \bibnamefont
  {Christianson}}, \bibinfo {author} {\bibfnamefont {A.~S.}\ \bibnamefont
  {Sefat}}, \bibinfo {author} {\bibfnamefont {B.~C.}\ \bibnamefont {Sales}},
  \bibinfo {author} {\bibfnamefont {M.~D.}\ \bibnamefont {Lumsden}}, \bibinfo
  {author} {\bibfnamefont {R.}~\bibnamefont {Jin}}, \bibinfo {author}
  {\bibfnamefont {E.~A.}\ \bibnamefont {Payzant}}, \bibinfo {author}
  {\bibfnamefont {D.}~\bibnamefont {Mandrus}}, \bibinfo {author} {\bibfnamefont
  {Y.}~\bibnamefont {Luan}}, \bibinfo {author} {\bibfnamefont {V.}~\bibnamefont
  {Keppens}}, \bibinfo {author} {\bibfnamefont {V.}~\bibnamefont
  {Varadarajan}}, \bibinfo {author} {\bibfnamefont {J.~W.}\ \bibnamefont
  {Brill}}, \bibinfo {author} {\bibfnamefont {R.~P.}\ \bibnamefont {Hermann}},
  \bibinfo {author} {\bibfnamefont {M.~T.}\ \bibnamefont {Sougrati}}, \bibinfo
  {author} {\bibfnamefont {F.}~\bibnamefont {Grandjean}}, \ and\ \bibinfo
  {author} {\bibfnamefont {G.~J.}\ \bibnamefont {Long}},\ }\href {\doibase
  10.1103/PhysRevB.78.094517} {\bibfield  {journal} {\bibinfo  {journal} {Phys.
  Rev. B}\ }\textbf {\bibinfo {volume} {78}},\ \bibinfo {pages} {094517}
  (\bibinfo {year} {2008})}\BibitemShut {NoStop}%
\bibitem [{\citenamefont {McGuire}\ \emph {et~al.}(2009)\citenamefont
  {McGuire}, \citenamefont {Hermann}, \citenamefont {Sefat}, \citenamefont
  {Sales}, \citenamefont {Jin}, \citenamefont {Mandrus}, \citenamefont
  {Grandjean},\ and\ \citenamefont {Long}}]{McGuire2009LnFeAsO}%
  \BibitemOpen
  \bibfield  {author} {\bibinfo {author} {\bibfnamefont {M.~A.}\ \bibnamefont
  {McGuire}}, \bibinfo {author} {\bibfnamefont {R.~P.}\ \bibnamefont
  {Hermann}}, \bibinfo {author} {\bibfnamefont {A.~S.}\ \bibnamefont {Sefat}},
  \bibinfo {author} {\bibfnamefont {B.~C.}\ \bibnamefont {Sales}}, \bibinfo
  {author} {\bibfnamefont {R.}~\bibnamefont {Jin}}, \bibinfo {author}
  {\bibfnamefont {D.}~\bibnamefont {Mandrus}}, \bibinfo {author} {\bibfnamefont
  {F.}~\bibnamefont {Grandjean}}, \ and\ \bibinfo {author} {\bibfnamefont
  {G.~J.}\ \bibnamefont {Long}},\ }\href {\doibase
  10.1088/1367-2630/11/2/025011} {\bibfield  {journal} {\bibinfo  {journal}
  {New J. Phys.}\ }\textbf {\bibinfo {volume} {11}},\ \bibinfo {pages} {025011}
  (\bibinfo {year} {2009})}\BibitemShut {NoStop}%
\bibitem [{\citenamefont {Tanatar}\ \emph {et~al.}(2009)\citenamefont
  {Tanatar}, \citenamefont {Ni}, \citenamefont {Samolyuk}, \citenamefont
  {Bud'ko}, \citenamefont {Canfield},\ and\ \citenamefont
  {Prozorov}}]{TanatarPRB2009}%
  \BibitemOpen
  \bibfield  {author} {\bibinfo {author} {\bibfnamefont {M.~A.}\ \bibnamefont
  {Tanatar}}, \bibinfo {author} {\bibfnamefont {N.}~\bibnamefont {Ni}},
  \bibinfo {author} {\bibfnamefont {G.~D.}\ \bibnamefont {Samolyuk}}, \bibinfo
  {author} {\bibfnamefont {S.~L.}\ \bibnamefont {Bud'ko}}, \bibinfo {author}
  {\bibfnamefont {P.~C.}\ \bibnamefont {Canfield}}, \ and\ \bibinfo {author}
  {\bibfnamefont {R.}~\bibnamefont {Prozorov}},\ }\href {\doibase
  10.1103/PhysRevB.79.134528} {\bibfield  {journal} {\bibinfo  {journal} {Phys.
  Rev. B}\ }\textbf {\bibinfo {volume} {79}},\ \bibinfo {pages} {134528}
  (\bibinfo {year} {2009})}\BibitemShut {NoStop}%
\bibitem [{\citenamefont {Carbotte}\ and\ \citenamefont
  {Schachinger}(2011)}]{Carbotte2011}%
  \BibitemOpen
  \bibfield  {author} {\bibinfo {author} {\bibfnamefont {J.~P.}\ \bibnamefont
  {Carbotte}}\ and\ \bibinfo {author} {\bibfnamefont {E.}~\bibnamefont
  {Schachinger}},\ }\href@noop {} {\bibfield  {journal} {\bibinfo  {journal}
  {J. Supercond. Nov. Magn.}\ }\textbf {\bibinfo {volume} {24}},\ \bibinfo
  {pages} {2269} (\bibinfo {year} {2011})}\BibitemShut {NoStop}%
\bibitem [{\citenamefont {Li}\ \emph {et~al.}(2010)\citenamefont {Li},
  \citenamefont {Broholm}, \citenamefont {Vaknin}, \citenamefont {Fernandes},
  \citenamefont {Abernathy}, \citenamefont {Stone}, \citenamefont {Pratt},
  \citenamefont {Tian}, \citenamefont {Qiu}, \citenamefont {Ni}, \citenamefont
  {Diallo}, \citenamefont {Zarestky}, \citenamefont {Bud'ko}, \citenamefont
  {Canfield},\ and\ \citenamefont {McQueeney}}]{LiPRB2010}%
  \BibitemOpen
  \bibfield  {author} {\bibinfo {author} {\bibfnamefont {H.-F.}\ \bibnamefont
  {Li}}, \bibinfo {author} {\bibfnamefont {C.}~\bibnamefont {Broholm}},
  \bibinfo {author} {\bibfnamefont {D.}~\bibnamefont {Vaknin}}, \bibinfo
  {author} {\bibfnamefont {R.~M.}\ \bibnamefont {Fernandes}}, \bibinfo {author}
  {\bibfnamefont {D.~L.}\ \bibnamefont {Abernathy}}, \bibinfo {author}
  {\bibfnamefont {M.~B.}\ \bibnamefont {Stone}}, \bibinfo {author}
  {\bibfnamefont {D.~K.}\ \bibnamefont {Pratt}}, \bibinfo {author}
  {\bibfnamefont {W.}~\bibnamefont {Tian}}, \bibinfo {author} {\bibfnamefont
  {Y.}~\bibnamefont {Qiu}}, \bibinfo {author} {\bibfnamefont {N.}~\bibnamefont
  {Ni}}, \bibinfo {author} {\bibfnamefont {S.~O.}\ \bibnamefont {Diallo}},
  \bibinfo {author} {\bibfnamefont {J.~L.}\ \bibnamefont {Zarestky}}, \bibinfo
  {author} {\bibfnamefont {S.~L.}\ \bibnamefont {Bud'ko}}, \bibinfo {author}
  {\bibfnamefont {P.~C.}\ \bibnamefont {Canfield}}, \ and\ \bibinfo {author}
  {\bibfnamefont {R.~J.}\ \bibnamefont {McQueeney}},\ }\href {\doibase
  10.1103/PhysRevB.82.140503} {\bibfield  {journal} {\bibinfo  {journal} {Phys.
  Rev. B}\ }\textbf {\bibinfo {volume} {82}},\ \bibinfo {pages} {140503}
  (\bibinfo {year} {2010})}\BibitemShut {NoStop}%
\bibitem [{\citenamefont {Pratt}\ \emph {et~al.}(2010)\citenamefont {Pratt},
  \citenamefont {Kreyssig}, \citenamefont {Nandi}, \citenamefont {Ni},
  \citenamefont {Thaler}, \citenamefont {Lumsden}, \citenamefont {Tian},
  \citenamefont {Zarestky}, \citenamefont {Bud'ko}, \citenamefont {Canfield},
  \citenamefont {Goldman},\ and\ \citenamefont {McQueeney}}]{PrattPRB2010}%
  \BibitemOpen
  \bibfield  {author} {\bibinfo {author} {\bibfnamefont {D.~K.}\ \bibnamefont
  {Pratt}}, \bibinfo {author} {\bibfnamefont {A.}~\bibnamefont {Kreyssig}},
  \bibinfo {author} {\bibfnamefont {S.}~\bibnamefont {Nandi}}, \bibinfo
  {author} {\bibfnamefont {N.}~\bibnamefont {Ni}}, \bibinfo {author}
  {\bibfnamefont {A.}~\bibnamefont {Thaler}}, \bibinfo {author} {\bibfnamefont
  {M.~D.}\ \bibnamefont {Lumsden}}, \bibinfo {author} {\bibfnamefont
  {W.}~\bibnamefont {Tian}}, \bibinfo {author} {\bibfnamefont {J.~L.}\
  \bibnamefont {Zarestky}}, \bibinfo {author} {\bibfnamefont {S.~L.}\
  \bibnamefont {Bud'ko}}, \bibinfo {author} {\bibfnamefont {P.~C.}\
  \bibnamefont {Canfield}}, \bibinfo {author} {\bibfnamefont {A.~I.}\
  \bibnamefont {Goldman}}, \ and\ \bibinfo {author} {\bibfnamefont {R.~J.}\
  \bibnamefont {McQueeney}},\ }\href {\doibase 10.1103/PhysRevB.81.140510}
  {\bibfield  {journal} {\bibinfo  {journal} {Phys. Rev. B}\ }\textbf {\bibinfo
  {volume} {81}},\ \bibinfo {pages} {140510} (\bibinfo {year}
  {2010})}\BibitemShut {NoStop}%
\bibitem [{\citenamefont {Fernandes}\ and\ \citenamefont
  {Schmalian}(2012)}]{Fernandes2012}%
  \BibitemOpen
  \bibfield  {author} {\bibinfo {author} {\bibfnamefont {R.~M.}\ \bibnamefont
  {Fernandes}}\ and\ \bibinfo {author} {\bibfnamefont {J.}~\bibnamefont
  {Schmalian}},\ }\href@noop {} {\bibfield  {journal} {\bibinfo  {journal}
  {Supercond. Sci. Technol.}\ }\textbf {\bibinfo {volume} {25}},\ \bibinfo
  {pages} {084005} (\bibinfo {year} {2012})}\BibitemShut {NoStop}%
\bibitem [{\citenamefont {Fernandes}\ \emph
  {et~al.}(2010{\natexlab{b}})\citenamefont {Fernandes}, \citenamefont
  {VanBebber}, \citenamefont {Bhattacharya}, \citenamefont {Chandra},
  \citenamefont {Keppens}, \citenamefont {Mandrus}, \citenamefont {McGuire},
  \citenamefont {Sales}, \citenamefont {Sefat},\ and\ \citenamefont
  {Schmalian}}]{FernandesPRL2010}%
  \BibitemOpen
  \bibfield  {author} {\bibinfo {author} {\bibfnamefont {R.~M.}\ \bibnamefont
  {Fernandes}}, \bibinfo {author} {\bibfnamefont {L.~H.}\ \bibnamefont
  {VanBebber}}, \bibinfo {author} {\bibfnamefont {S.}~\bibnamefont
  {Bhattacharya}}, \bibinfo {author} {\bibfnamefont {P.}~\bibnamefont
  {Chandra}}, \bibinfo {author} {\bibfnamefont {V.}~\bibnamefont {Keppens}},
  \bibinfo {author} {\bibfnamefont {D.}~\bibnamefont {Mandrus}}, \bibinfo
  {author} {\bibfnamefont {M.~A.}\ \bibnamefont {McGuire}}, \bibinfo {author}
  {\bibfnamefont {B.~C.}\ \bibnamefont {Sales}}, \bibinfo {author}
  {\bibfnamefont {A.~S.}\ \bibnamefont {Sefat}}, \ and\ \bibinfo {author}
  {\bibfnamefont {J.}~\bibnamefont {Schmalian}},\ }\href {\doibase
  10.1103/PhysRevLett.105.157003} {\bibfield  {journal} {\bibinfo  {journal}
  {Phys. Rev. Lett.}\ }\textbf {\bibinfo {volume} {105}},\ \bibinfo {pages}
  {157003} (\bibinfo {year} {2010}{\natexlab{b}})}\BibitemShut {NoStop}%
\bibitem [{\citenamefont {Chu}\ \emph {et~al.}(2010)\citenamefont {Chu},
  \citenamefont {Analytis}, \citenamefont {De~Greve}, \citenamefont {McMahon},
  \citenamefont {Islam}, \citenamefont {Yamamoto},\ and\ \citenamefont
  {Fisher}}]{ChuScience2010}%
  \BibitemOpen
  \bibfield  {author} {\bibinfo {author} {\bibfnamefont {J.-H.}\ \bibnamefont
  {Chu}}, \bibinfo {author} {\bibfnamefont {J.~G.}\ \bibnamefont {Analytis}},
  \bibinfo {author} {\bibfnamefont {K.}~\bibnamefont {De~Greve}}, \bibinfo
  {author} {\bibfnamefont {P.~L.}\ \bibnamefont {McMahon}}, \bibinfo {author}
  {\bibfnamefont {Z.}~\bibnamefont {Islam}}, \bibinfo {author} {\bibfnamefont
  {Y.}~\bibnamefont {Yamamoto}}, \ and\ \bibinfo {author} {\bibfnamefont
  {I.~R.}\ \bibnamefont {Fisher}},\ }\href@noop {} {\bibfield  {journal}
  {\bibinfo  {journal} {Science}\ }\textbf {\bibinfo {volume} {329}},\ \bibinfo
  {pages} {824} (\bibinfo {year} {2010})}\BibitemShut {NoStop}%
\bibitem [{\citenamefont {Chu}\ \emph {et~al.}(2012)\citenamefont {Chu},
  \citenamefont {Kuo}, \citenamefont {Analytis},\ and\ \citenamefont
  {Fisher}}]{ChuScience2012}%
  \BibitemOpen
  \bibfield  {author} {\bibinfo {author} {\bibfnamefont {J.-H.}\ \bibnamefont
  {Chu}}, \bibinfo {author} {\bibfnamefont {H.-H.}\ \bibnamefont {Kuo}},
  \bibinfo {author} {\bibfnamefont {J.~G.}\ \bibnamefont {Analytis}}, \ and\
  \bibinfo {author} {\bibfnamefont {I.~R.}\ \bibnamefont {Fisher}},\
  }\href@noop {} {\bibfield  {journal} {\bibinfo  {journal} {Science}\ }\textbf
  {\bibinfo {volume} {337}},\ \bibinfo {pages} {710} (\bibinfo {year}
  {2012})}\BibitemShut {NoStop}%
\end{thebibliography}

%

\end{document}